\documentclass[aps,pre,onecolumn,showpacs,superscriptaddress,groupedaddress]{revtex4}
\usepackage[version=3]{mhchem} 

\usepackage[]{hyperref}
\usepackage{graphicx}
\usepackage{amsmath}
\usepackage[normalem]{ulem}
\usepackage{siunitx}
\usepackage{soul}
\usepackage{amssymb}
\makeatletter
\usepackage{rotating}
\usepackage{url}
\usepackage{anyfontsize}
\usepackage{blindtext}
\usepackage[utf8]{inputenc}
\usepackage{xr}
\newcommand*{\rom}[1]{\expandafter\@slowromancap\romannumeral #1@}
\makeatother
\usepackage{lipsum}
\usepackage{xcolor}
\usepackage{setspace}
\usepackage{ccaption}

\usepackage{floatrow}
\usepackage[font= large,label font=bf,labelformat=simple]{subfig}

\usepackage[T1]{fontenc}
\usepackage{lipsum}
\usepackage{tabularx}

\begin{document}

\title[An \textsf{achemso} demo]
  {Statistical Mechanical theory for spatio-temporal evolution of Intra-tumor heterogeneity in cancers: Analysis of Multiregion sequencing data}

\author{Sumit Sinha}
\affiliation{Department of Physics, University of Texas at Austin, Austin, TX 78712, USA.}
\author{Xin Li}
\affiliation{Department of Chemistry, University of Texas at Austin, Austin, TX 78712, USA.}
\author{D. Thirumalai}
\email{dave.thirumalai@gmail.com*}
\affiliation{Department of Chemistry, University of Texas at Austin, Austin, TX 78712, USA.}

\singlespacing

\begin{abstract}
\textcolor{black}{Variations in characteristics from one region (sub-population) to another is commonly observed in complex systems, such as glasses and a collection of cells. Such variations are  manifestations of heterogeneity, whose spatial and temporal behavior is hard to describe theoretically. In the context of cancer, intra-tumor heterogeneity (ITH), characterized by cells with genetic and phenotypic variability that co-exist within a single tumor, is often the cause of ineffective therapy and recurrence of cancer. Next-generation sequencing, obtained by sampling multiple regions of a single tumor (multi-region sequencing, M-Seq), has vividly demonstrated the pervasive nature of ITH, raising the need for a theory that accounts for  evolution of tumor heterogeneity. Here, we develop a statistical mechanical theory to quantify ITH, using the Hamming distance, between genetic mutations in distinct regions within a single tumor. An analytic expression for ITH, expressed in terms of cell division probability ($\alpha$) and mutation probability ($p$), is validated using cellular-automaton type simulations. Application of the theory successfully captures ITH extracted from M-seq data in patients with exogenous cancers (melanoma and lung). The theory, based on punctuated evolution at the early stages of the tumor followed by neutral evolution, is accurate provided the spatial variation in the tumor mutation burden is not large. We show that there are substantial variations in ITH in distinct regions of a single solid tumor, which supports the notion that distinct subclones could co-exist. The simulations show that there are substantial variations in the sub-populations, with the ITH increasing as the distance between the regions increases. The analytical and simulation framework developed here could be used in the quantitative analyses of the experimental (M-Seq) data. More broadly, our theory is likely to be useful in analyzing dynamic heterogeneity in complex systems such as super-cooled liquids.}
\end{abstract}

\date{\today}
\maketitle


\section{ Introduction}
Heterogeneity is pervasive on all scales in abiotic and biological systems. In glass forming systems, the dynamics in a seemingly homogeneous sample shows that slow and fast moving particles coexist within one sample, a phenomena referred to as dynamical heterogeneity \cite{berthier2011theoretical, kirkpatrick2015colloquium}. As a consequence, there are variations in  properties (distribution of energies for example) in different regions within a single sample of a supercooled liquid near the glass transition \cite{kirkpatrick2015colloquium, thirumalai1989ergodic}. In biology, it is becoming increasingly clear that heterogeneity dominates on all length scales \cite{altschuler2010cellular}, from molecules to chromosome conformations, and in the behavior of different cells. Nowhere is the heterogeneous behavior more relevant than in cancers in which there are genotypic and phenotypic variations among different sub-populations within a single tumor, which is commonly referred to as intratumor heterogeneity (ITH) \cite{zahir2020characterizing, caswell2021molecular, mcgranahan2017clonal, heppner1984tumor}.  

\textcolor{black}{Cancer is an evolutionary disease that is, in most cases, triggered through accumulation of harmful mutations over a period of several decades \cite{merlo2006cancer}. Both exogenous (external environmental factors) and endogenous factors (such as DNA replication errors) \cite{tomasetti2017stem, tomasetti2015variation} give rise to harmful mutations. Thus, cancers may be classified as, ``exogenous" (melanoma, lung etc.) and ``endogenous" (kidney, brain etc.) based on the origin of mutations. With the advent of next-generation sequencing technologies, one can characterize the genetic information of cancer cells at unprecedented whole-genome level resolution \cite{manolio2010genomewide}. A few hundreds of ``driver mutations", which bestow significant growth or fitness advantage to the cancer cells, have been discovered in different cancers through cancer genome projects \cite{vogelstein2013cancer}. However, the evolutionary dynamics of cancers, which is encoded in the mutational history, is still elusive \cite{ciriello2021many}. } In particular, a quantitative description of ITH, which could be readily used to analyze experimental data, does not exist. 

\textcolor{black}{In the linear evolution model for tumor growth, proposed in a seminal work \cite{nowell1976clonal}, cells accumulate driver mutations sequentially, which outcompete the preexisting clonal population through selective sweeps. However, most cancers exhibit extreme heterogeneity, as cells with distinct genetic and phenotypic characteristics, co-exist within a single tumor\cite{marusyk2010tumor, hinohara2019intratumoral, michor2010origins}. This finding is in apparent contrast to the linear evolution model. As an alternative, branched evolution in which multiple subclones grow simultaneously in the same tumor, with rare selective sweeps, is supported by many next-generation sequencing experiments \cite{gerlinger2012intratumor, gerlinger2014, swanton2018cancer}. Besides, neutral evolution, marked by lack of selection or fitness advantage, has been introduced for several cancer types\cite{williams2016identification,sottoriva2015big, sun2018big}. Recently, a punctuated evolution\cite{field2018punctuated} model was invoked for some cancers. Instead of accumulating mutations sequentially and gradually, as in the previous models, a burst of mutations may occur in a short period of time in the early stage of tumor evolution. Although there are several models for cancer evolution \cite{davis2017tumor,li2020cooperation}, it is universally agreed that intra-tumor heterogeneity (ITH) is common among different cancers. Pervasive ITH is one of the leading reasons for cancer resistance to conventional therapy\cite{burrell2013causes}. Hence, the study of ITH is of utmost importance both as a fundamental problem in the physics of cancers and for cancer treatment. }

Due to the spatial genetic variations among cancer cells, traditional single biopsies greatly underestimate the ITH in cancer patients \cite{swanton2012intratumor, gerlinger2014,michor2010origins}. Lack of quantitative understanding of the spatio-temporal variation in ITH is often blamed for ineffective therapeutic strategies as they target only part of tumor cell populations \cite{gerlinger2012intratumor, gerlinger2014, morrissy2017spatial}. Although known for decades, it is possibly only through Multiregion sequencing (M-Seq) data has it became abundantly clear that ITH is pervasive in many cancers. In M-Seq experiments, multiple regions in a single tumor are sequenced \cite{gerlinger2012intratumor, gerlinger2014}. These experiments for several cancers and many patients reveal that there are substantial region-to-region diversity in the mutated genes. Therefore, any theory for ITH must take spatial variations into account \cite{swanton2012intratumor,gonzalez2002metapopulation}. The implications of the M-Seq data analyses serves as the main inspiration for the present theoretical study.

Previously, several theoretical studies estimate ITH of cancer patients from sequencing data \cite{iwasa2011evolutionary,durrett2011intratumor, gonzalez2002metapopulation, thalhauser2010selection,martens2011spatial,anderson2006tumor,waclaw2015spatial, paterson2016exactly, antal2015spatial, werner2020measuring}. However, these studies have limitations, preventing their use in estimating ITH from M-Seq experiments. Patient's ITH is often evaluated using the Simpson's index \cite{durrett2011intratumor,iwasa2011evolutionary} where a well-mixed (spatially homogeneous) tumor evolution model is considered, which is clearly not suitable for analyzing the M-Seq data. In a pioneering study Gonazalez-Garcia et. al. \cite{gonzalez2002metapopulation}, which took into account spatial variations, depicted the \textit{in silico} results as a two-dimensional map, like the one in Fig. 1. Because they investigated the frequency of cell populations with mutations in only two genes, it limits the applicability of the model to M-Seq data. In a promising study, a spatial computational tumor model was created \cite{waclaw2015spatial}, which showed that shared genetic alterations decrease as a function of inter-cellular distance. However, they \cite{waclaw2015spatial} did not draw firm conclusions on ITH. In particular, they did not compare the available data from M-Seq experiments and the predictions from the computational model. 
 
Our work is closest in spirit to the seminal studies by Curtis and coworkers, who used an Approximate Bayesian Computation methodology\cite{sottoriva2015big} to propose the Big Bang dynamics, first in the context of  human colorectal cancer. They established that, after initially acquiring mutations, the tumor progresses without stringent selection.  Subsequently, they \cite{sun2017between} created a spatial tumor growth simulation model to predict the heterogeneity determined  in the multi-region sequencing experiments. Although the initial studies focused on colorectal cancer, it was later realized that effective neutrality (tumor growth without strict selection) is likely to be more generally valid \cite{sun2018big} for other cancers as well. Remarkable as these studies are, to the best of our knowledge, no prior study has developed a theoretical framework to calculate ITH analytically, so that it can be readily used to analyze M-Seq experiments. We fill this missing gap in this work by creating a statistical mechanical theory complemented by simulations.
 
To quantitatively estimate ITH revealed in M-Seq experiments, we develop a theory applicable to solid tumors based on neutral evolution of tumor in three dimensions inspired by recent studies \cite{sun2018big, sottoriva2015big}. We note parenthetically that although there is considerable debate on the nature of tumor evolution (neutral or with selection, linear and punctuated), it is universally accepted that ITH is pervasive in many cancers. Our theory has two parameters: \textcolor{black}{(i) Cell division probability ($\alpha$), and (ii) Mutation probability ($p$).} The central results of the study are: (\rom{1}) \textcolor{black}{A closed form expression for ITH is obtained, which is written as, $ITH(\alpha, p, t)=G(p, t)F(\alpha)$, where $G(p,t)$ is the ITH with $\alpha=1$. The function $F(\alpha)$ is the scale factor that accounts for the possibility that $\alpha$ is less than unity ($0.5<\alpha<1$). (\rom{2}) $G(p,t)$} is determined by the correlations induced in cell lineages that arise due to cell division. The magnitude of $G(p, t)$ depends on the geometry in which the tissue is embedded. (\rom{3}) The scale factor, $F(\alpha)$, increases linearly upon increasing $\alpha$, physically implying that ITH decreases on increasing apoptosis (i.e reducing cell division probability). (\rom{4}) The utility of our theory, quantified using Pearson correlation ($\rho$), is established by directly comparing with the M-Seq data for cancers originating in four tissue types-Skin \cite{harbst2016} (8 patients), Lung \cite{debruin2014, jamal2017tracking} (60 patients), and Kidney \cite{gerlinger2014} (10 patients). We obtained $\rho=0.97$ for skin, $\rho=0.87$ for lung and $\rho=0.51$ for Kidney, indicative that our theory can capture M-Seq experiments for exogenous cancers (i.e skin and lung). \textcolor{black}{(\rom{5}) We find that the effectiveness of our theory in capturing experiments is related to the spatial distribution of tumor mutation burden (TMB), quantified using the coefficient of TMB variation ($c_v$).} Exogenous cancers have smaller $c_v$ (similar to predictions from theory) compared to endogenous cancer and hence the theory captures the ITH observed cancers originating in skin and lung. \textcolor{black}{The theory provides a testable framework for analyzing ITH extracted from M-Seq experiments. (\rom{6}) Besides confirming the validity of the theory, the simulations vividly illustrate ITH as manifestation of massive variations in the Hamming distance from one region to another. This picture, that emerges from the simulations, is in accord with the M-Seq experiments.   }   

 \section{ Extraction of ITH from M-Seq data}
 
 \subsection{Analysis of Experimental data}
 \label{method_expt}
\begin{figure}
\includegraphics[clip,width=1\textwidth]{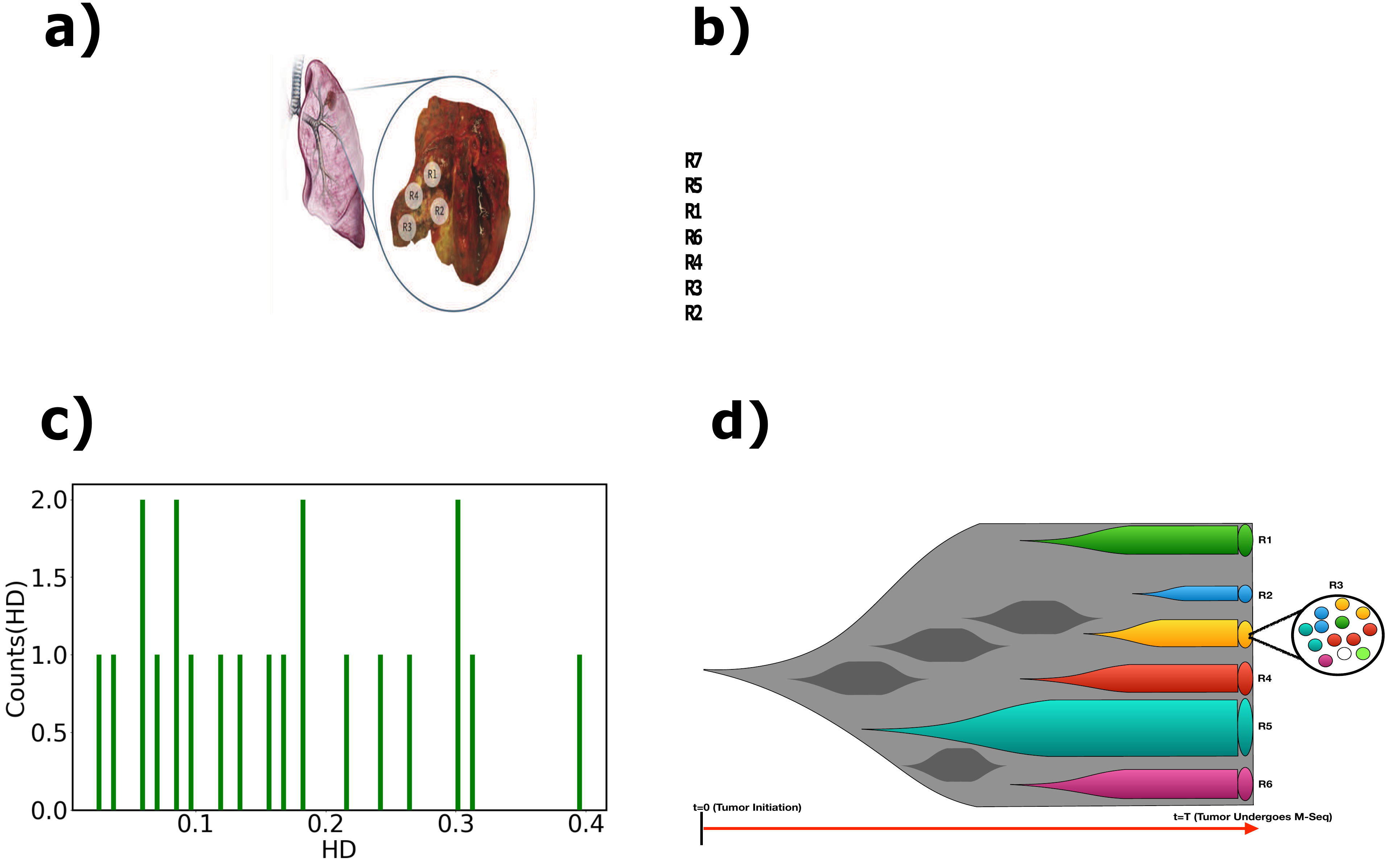}
\caption{{\bf Schematic of Multiregion Sequencing (M-Seq)} {\bf (a)} A tumor sample (dark red) located in the lung (right lobe) of a cancer patient \cite{jamal2017tracking}. An enlarged view, depicted in the oval, is on the right. The labels  R1, R2, R3 and R4 are the regions where the cells were extracted by biopsy and whole exome sequencing, referred to as Multiregion Sequencing (M-Seq), is performed.
~{\bf (b)} A typical readout of the M-Seq experiment \cite{gerlinger2014} for a kidney cancer patient where M-Seq in seven regions were conducted. In this heat map representation, the presence of mutated (non-mutated) gene in a region is denoted as a yellow (blue) box. The number of columns represent the total number of uniquely mutated genes ($n$) in the entire tumor ($n=83$ in this case). In the row adjacent to the heat map, \textcolor{black}{the mutated genes are listed and the probable driver mutations are displayed in magenta}. The rows represents the seven regions that were sampled.{\bf (c)} Distribution of the Hamming Distance (HD) for the M-Seq data in (b). The y-axis (x-axis) shows the counts (HD values). Sum of the counts $=C_{2}^{7}=21$, the total number of region pairs for M-Seq data in (b). {\bf (d)} A schematic for neutral evolution for cancer. The \textcolor{black}{regions R1 to R6 with different colors represent the distinct genetic composition measured using M-Seq.} The schematic represents neutral evolution because the size of different regions is proportional to their lifetime, cyan (blue) region being the biggest. We zoomed in on R3 to show the distinct genetic makeup of the cells. The \textcolor{black}{dark grey} lineages are do not survive during the course of tumor evolution. }
\label{fig:MSeq_figure}
\end{figure}

 We first analyzed the experimental data because it sets the stage for the development of the theoretical model. We used the M-Seq data-sets that were collected for cancers in four tissue types -Skin \cite{harbst2016}, Lung \cite{debruin2014, jamal2017tracking}, Esophagus \cite{cao2015} and Kidney \cite{gerlinger2014}. M-Seq \cite{gerlinger2012intratumor, gerlinger2014, harbst2016, debruin2014, zhang2014intratumor, cao2015} refers to whole exome sequencing of multiple regions in a cancerous tumor (Figure \ref{fig:MSeq_figure}a). 
 A typical readout of a M-Seq data is represented as a matrix (see Figure~\ref{fig:MSeq_figure}b), where the non-synonymous mutations detected across all the regions are displayed in the form of a binary heat map. \textcolor{black}{The string (a one dimensional array that contains the list of genes) of length, $n$, in the M-Seq data is the total number of uniquely mutated genes in the whole tumor. In principle, the blueprint of cancer evolution over time scales spanning decades is contained in these readouts, \cite{williams2019measuring} which is illustrated schematically in Figure \ref{fig:MSeq_figure}d.} Depending on the pattern of the heat-maps, a measure that quantifies ITH may be calculated for each readout. The average ITH values (explained below) are unique to each patient. We used Hamming distance (HD) as a measure of ITH. 
  
 We collected M-Seq readouts from four cancer types. Three of them (skin, lung and esophagus cancers) arise predominantly due to exogenous factors, while kidney cancer is caused by endogenous factors \cite{tomasetti2017stem}. We analyzed eight, sixty, two and ten publically available patient data for skin \cite{harbst2016}, lung \cite{debruin2014, jamal2017tracking}, esophageal \cite{cao2015} and kidney \cite{gerlinger2014}  cancers, respectively.
 
 
 {\bf ITH from experimental readouts:} The M-Seq data, represented as a matrix whose elements represent the biopsied regions (columns in Figure 1b) and the sequenced genes (rows in Figure 1b), generate averages over the cells in a specific region. We use Figure \ref{fig:MSeq_figure}b to illustrate how ITH is extracted from M-seq data. There are seven regions, labeled R1, R2, R3, R4, R5, R6 ,and R7. Let each column be $X^{R_i}$, where $i \in [1,2,3,4,5,6,7]$. The length\textcolor{black}{, $n$, of each row ($|X^{R_i}|$) gives the number of genes, which is 83 in Figure \ref{fig:MSeq_figure}b.} To quantify the average heterogeneity associated with the data in Figure \ref{fig:MSeq_figure}b, we calculated the Hamming distance (HD) \cite{zhai2017spatial, jiang2014statistical, martens2011spatial}, between all the pairs in the seven regions. \textcolor{black}{The total number of distinct pairs, denoted by $M$, is $C^{M}_2=C_{2}^{7}=21$ in Figure~\ref{fig:MSeq_figure}b}. Hamming distance between columns belonging to two regions, $A$ and $B$, is given by, $HD(X^{R_A}, X^{R_B})=\frac{1}{n}\sum_{i=1}^{n}|X^{R_A}_i-X^{R_B}_i|$, where $|..|$ is the absolute value. Note that $0\leq HD(X^{R_A}, X^{R_B})\leq 1$ with $0$ being homogeneous (no spatial variation), and $HD(X^{R_A}, X^{R_B})=1$ implies that the tumor is maximally heterogeneous. Table 1 in the SI shows the Hamming distances between the 21 pairs. After calculating the HD among all the possible pairs, we determined \textcolor{black}{the average value <HD>} (0.17 in Figure \ref{fig:MSeq_figure}b). Figure 1c shows the distribution of $HD$ between all the $21$ pairs. The $HD$ values range from $\approx 0.02$ to $\approx 0.40$, with a mean of $0.17$. The data, as presented, already shows that there are variations at the mutation level between different regions in a single tumor. We adopt the same \textcolor{black}{procedure to} calculate the ITH, measured in terms of the HD, for each patient from the M-Seq readout for the other cancer types (see the SI figures S2-S6 for details).

\subsection{Theory for ITH with neutral evolution}
\label{Theory_Model}
The theoretical model is based on generating evolutionary trajectories through space and time using 3D lattice representation of the tumor \textcolor{black}(illustrated in  Figure~S11 in the SI). Besides occupying a site on the lattice, \textcolor{black}{each cell $i$ carries genetic information in the form of a string $X^{i}$ whose length, $n$, is assumed to be a constant}. This is similar to the M-seq experiments (see Figure \ref{fig:MSeq_figure}b with $n=83$). Each site in the string represents a gene in the DNA. A gene that is not mutated is denoted by 0 while a mutated gene is represented by 1, which is the binary representation adopted in the M-seq experiments. \textcolor{black}{We initialize the tumor evolution by placing a single cell with $X^{i}=0$ at the center of the 3D lattice. The tumor evolves and generates an evolutionary trajectory in space and time according to the following rules.}


 {\bf (i) Cell division and replication:} At each time $t$, a cell could divide and give birth to a daughter cell, with probability, $\alpha$, provided \textcolor{black}{one of the 26 neighboring sites} in the 3D lattice is vacant.  \textcolor{black}{We are considering a growing tumor through punctuated evolution \cite{field2018punctuated} initially, with a few driver mutations, which implies that $\alpha>0.5$.} After cell division, the genetic information of the parent (P) cell is copied to the daughter ($X^{D}=X^{P}$). This biologically realistic copying mechanism is an essential part of the model as it leads to correlations in space and time \cite{hormoz2015inferring, hormoz2016inferring}. 

 {\bf (ii) Mutations:} During each time-step, a cell could acquire mutations at non-mutated sites with probability, $p$. Acquisition of mutation is independent of cell division, which is the case for exogenous mutations. For instance, let us denote the cell by index $m$, non-mutated site by $i$ and the string by $X^{m}_i$. The above statement can be represented mathematically as, $\mathbb{P}[X^{m}_{i}(t+1)=1 | X^{m}_{i}(t)=0 ]=p  $. It follows that the probability that a site is not mutated at $t+1$, given that it was not mutated at time $t$, is $1-p$, which we write as \textcolor{black}{ $\mathbb{P}[X^{m}_{i}(t+1)=0 | X^{m}_{i}(t)=0]=1-p$}. 
 
 {\bf (iii) Mutations are neutral:} \textcolor{black}{The mutations do not add fitness or selective advantages to cells since we only consider neutral evolution \cite{davis2017tumor} at later stages of tumor growth although initially we used punctuated evolutionary process.} Preliminary results using evolution with selection that confers fitness advantage to cancer cells yields qualitatively similar results (see the Figure S12 in the Supplementary Information (SI) ).

{\bf (iv) Mutations are irreversible:} We also assume that a site (gene in the M-Seq data), on string $X^{m}$, once mutated cannot revert to the original status in the future, \textcolor{black}{as is assumed in other studies \cite{el2015reconstruction}}. This can be rationalized by the following argument. In M-Seq experiments, the sites in the strings $X^m$ code for genes. Suppose we consider a gene, Q, with length $L$ base-pairs (bp), then the ratio of reverse mutation to forward mutation probabilities is negligible. Assuming that only a single nucleotide in gene Q is mutated, the reverse mutation probability ($\mathcal{P}_b$) is $\textcolor{black}{\mu_n}$, where $\textcolor{black}{\mu_n}$ is the mutation probability for a nucleotide. Note that mutation probability of a single nucleotide $\textcolor{black}{\mu_n}$ is different from mutation probability of gene, $p$. Similarly, the forward probability ($\mathcal{P}_f$) is $(L-1)\textcolor{black}{\mu_n}$. Therefore, $\frac{\mathcal{P}_b}{\mathcal{P}_f}=\frac{1}{L-1}\rightarrow 0$, because the average gene length $L\approx 10-15 ~kbp$ \cite{strachan1999human,grishkevich2014gene}. Therefore, once a nucleotide in the gene is mutated, the probability of reversing the mutation at that specific nucleotide is negligible. Mathematically, the irreversibility of mutations is written as, $\mathbb{P}[X^{m}_{i}(t+1)=0 | X^{m}_{i}(t)=1]=0$.

{\bf (v) Apoptosis:} At any time-step, a cell may undergo apoptosis with probability $1-\alpha$. Given the five rules, our goal is to calculate the average ITH, expressed as the mean HD, in a tumor containing $N$ cells. There are only two parameters in the model, cell division probability ($\alpha$), and mutation probability ($p$). Let $t$ be the time for the tumor to grow $N$ cells. We define \textcolor{black}{the $ITH(t)$ of the tumor} as,
\begin{equation}
 ITH(t) = \frac{2}{N(N-1)}\sum_{i=1}^{N}\sum_{j=i+1}^{N} HD[X^{i}(t),X^{j}(t)],
\label{eq:one}
\end{equation} 
where $HD[X^{i}(t_N),X^{j}(t_N)]$  is the Hamming distance (HD) between the DNA strings in cells $i$ and $j$ at time $t$. The method used to compute HD theoretically is similar to the one we employed to analyze M-seq experiments (see SI Table 1).

\subsection{Stochastic simulations for tumor growth}
To validate the theoretical results, we developed a three-dimensional lattice model of tumor evolution (see the details in section \rom{3} of the SI.) The cellular automaton based simulations mimic the theoretical rules for tumor evolution, which are described in the previous section. For all the simulation results, we used the string length, $n=300$, which is comparable to the string length in the M-Seq experiments.

\section{Results}
\subsection{ITH under neutral evolution}
In terms of cell division probability ($\alpha$) and mutation probability ($p$), \textcolor{black}{the $ITH(\alpha,p,t)$ of a neutrally evolving tumor in three dimensions (3D) may be expressed as,}
\begin{equation}
    \textcolor{black}{ITH(\alpha, p, t)\equiv G_{3D}(p,t)F(\alpha)=2x_t(1-x_t)(0.2\alpha+0.77).}
    \label{ansatz_ith}
\end{equation}
In Eq.\ref{ansatz_ith}, \textcolor{black}{$G_{3D}(p,t) =2x_t(1-x_t)$ is the value of ITH when $\alpha=1$, $x_t=(1-p)^t$}, and $F(\alpha)$ is a scale factor, which ensures that $\alpha$ is less than unity ($0.5<\alpha<1$).  \textcolor{black}{This is the central result in this work.} The seemingly simple analytic expression for $ITH(\alpha, p, t)$ in Eq. \ref{ansatz_ith} shows that it factorizes into a product of the two functions one of which depends solely on $p$ and the other monotonically increases with $\alpha$.


{\bf Branches, correlated and uncorrelated evolution:} Let us first provide insights into Eq. \ref{eq:one} for tumor evolution with $\alpha=1$. Figure \ref{fig:th_g}a illustrates a sample evolutionary tree at $t=0, 1, 2, 3$ for $\alpha=1$. Figure S15 shows the snapshots of tumor evolution for $\alpha=1$ obtained from lattice simulations. In this case, the growth process may be understood in terms of a directed tree, composed of cells and  directed edges. Because $\alpha=1$, once a cell or an edge is created it remains throughout the evolutionary process. The edges provide the child-parent relationship between two cells. For instance, in Figure \ref{fig:th_g}a, the edges from $1\rightarrow2$ means that cell $2$ was born from cell $1$. The evolutionary tree, which is an imprint of the trajectories, has many branches. A branch is a unique path traversed from the origin (cell $1$) to the last cell (no directed edge from a node) following the directed edges. In Figure \ref{fig:th_g}a, $1\rightarrow2\rightarrow3\rightarrow5$ is
the longest branch at $t = 3$. In 3D, these branches are curved implying that the Euclidean distance from cell $1$ to the end of the branch (the corresponding leaf node), is not equal to the number of nodes in the branch. However, because cell-division occurs only if there is a vacant neighbor, we assume that the curvature associated with the branches are negligible, rendering them as linear. With this assumption, a 3D evolutionary tree may be pictured as consisting of many linear branches. The linear branches correspond to the evolution of cells on a semi-infinite 1D lattice where growth occurs unidirectionally away from the origin. For this reason it is instructive to calculate ITH for a 1D-semi infinite lattice. For completeness, we also estimate the number of branches in all the geometries - 1D semi-infinite, 1D infinite, 2D and 3D evolutionary trees. Let the number of branches in an evolutionary tree be $N_{br}$. For semi-infinite lattice $N_{br}=1$, for infinite 1D lattice $N_{br}=2$, for a 2D evolutionary tree $N_{br}\sim R$, and for a 3D evolutionary tree $N_{br}\sim R^2$. Here, $R$ is the radius of the tumor in two and three dimensions. In $d$ dimensions, $N_{br}\sim R^{d-1}.$

The heterogeneity between the cell pairs depends on whether they belong to the same branch or not. Pairs of cells sharing a common branch will have correlations in their genetic information whereas cells that do not share any common branch segment will evolve independently. For instance, in Figure \ref{fig:th_g}a, the evolution of node $8$ is independent of node $5$ as they belong to different branches. On the other hand, the evolution of node $5$ is correlated with evolution of node $2$ because they are in the same branch. Figures \ref{fig:th_g}b and \ref{fig:th_g}c represent the schematic for cells undergoing independent and correlated evolution. To evaluate $G_{3D}(p, t)$, we first calculate heterogeneity among cells that undergo independent and correlated evolution.

\textit{{\bf Heterogeneity in independently evolving cells:}} Figure \ref{fig:th_g}b shows a schematic of cell pairs, labelled as 1 and 2, evolving independently in time. The genes in their respective DNA, $X^1$ and $X^2$, are not mutated at time $t=0$ ( $X^1_i=X^{2}_i=0$ for $1\leq i\leq n$). Given the rules for generating the evolution trajectories, presented in section \rom{2}.B, we evaluate the average heterogeneity among a pair of cells evolving independently, $\langle HD_{ind}[X^1(t),X^2(t)] \rangle$. Here, $\langle ... \rangle$ represents the ensemble average over all the cell pairs (see SI for an additional explanation).

It is shown in the Appendix that the average heterogeneity between a pair of cells evolving independently ($\langle HD_{ind}[X^1(t),X^2(t)] \rangle$) is,
\begin{equation}
\langle HD_{ind}[X^1(t),X^2(t)]  \rangle=2x_{t}(1-x_{t}),
\label{heteroindependent_here}
\end{equation}
where $x_{t}=(1-p)^{t}$ is the probability that a gene is not mutated till time $t$.

\textit{{\bf Heterogeneity in cells with correlated evolution:}} Figure \ref{fig:th_g}c shows a schematic of an evolutionary trajectory for cell pairs, labelled as 1 and 2, that is correlated in time. During correlated evolution, cell pairs have a common ancestor till $t>0$. Let us consider the case when a cell evolves acquiring mutations from $t=0$ till time $t$. After time $t$, it gives birth to cell $2$. During the birth process, the content in string $X^1$ is copied to string $X^2$, which introduces correlations between $X^1$ and $X^2$ . We wish to evaluate $\langle HD_{corr}[X^1(t+s),X^2(t+s)] \rangle$ between pairs of cells that undergo correlated evolution at some later time $t+s$. The average heterogeneity ($\langle HD_{corr}[X^1(t),X^2(t)] \rangle$) in this case is (see Appendix for details),

\begin{equation}
\langle HD_{corr}[X^1(t+s),X^2(t+s)]  \rangle=2x_{t+s}(1-x_{s}).
\label{correlated_t}
\end{equation}
 The expression in Eq. \ref{correlated_t}, gives the average heterogeneity between two cells which undergo a common evolution till $t$ but subsequently evolve independently till $t+s$. If we set $t=0$ in Eq.\ref{correlated_t}, we obtain the result for independent evolution in Eq.\ref{heteroindependent_here}. The result in Eq. \ref{correlated_t} is particularly interesting because it shows that the two cells are correlated in time because of the copying mechanism (daughter cell inherits all the genetic information in the mother cell) during cell-division. It has previously been noted that cell division induced correlations among cells share a common lineage \cite{hormoz2015inferring,hormoz2016inferring}. Because of correlated evolution, cell pairs that undergo correlated evolution have smaller heterogeneity compared to cells pairs that evolve independently ($\frac{HD_{ind}[X^1(t+s),X^2(t+s)]}{HD_{corr}[X^1(t+s),X^2(t+s)]}=\frac{1-x_{t+s}}{1-x_{s}}>1$).

{\bf Role of Branches in an evolutionary tree:} Having calculated the heterogeneity among cells evolving independently and in a correlated manner, we can evaluate the average heterogeneity within the 3D tumor at time $t$ for $\alpha=1$, $G_{3D}(t,p)$. The expression for $G_{3D}(t,p)$ is given as, 

\begin{equation}
 G_{3D}(t,p)=\frac{\int _1^R\int _1^{r_1}r_1^2r_2^2 F_{3D}(r_1,r_2,t)dr_2dr_1}{\int _1^R\int _1^{r_1}r_1^2r_2^2dr_2dr_1},
\label{ith_expr1}
\end{equation}
where $R$ is the radius of the tumor, and $F_{3D}(r_1,r_2,t)$ is the heterogeneity for a pair of cells at distances $r_1$ and $r_2$ from the origin at time $t$. Without loss of generality, we assume that $r_1 \geq r_2$. The general expression of $F_{3D}(r_1,r_2,t)$ may be written as, 
\begin{equation}
F_{3D}(r_1,r_2,t)=P_{br, 3D}\{2x_t(1-x_{t-r_2})\}+(1-P_{br, 3D})\{2x_t(1-x_t)\}
\label{f_expr}
\end{equation}
where $P_{br, 3D}$ denotes the probability that both the cells belong to the same branch of the 3D evolutionary tree. In 3D, $P_{br, 3D}=\frac{1}{\frac{4}{3}\pi r_1^2}$. Note that to obtain the expression of $P_{br, 3D}$, we assumed that the branches are linear. On evaluating the integrals in Eq. \ref{ith_expr1}, we obtain an analytical expression for $G_{3D}(p,t)$ given by,
\begin{multline}
G_{3D}(p,t)=\\
\frac{3 (1-p)^t}{\pi  \left(R^3-1\right)^2}\bigg(-\frac{2}{3} \pi  \left(R^3-1\right)^2 \big[(1-p)^t-1\big]+\frac{3}{4} (R-1)^2 (1-p)^t \left(R^2+2R+3\right) \\
-\frac{(1-p)^{t-R-1}}{\log ^4(1-p)}\bigg[9 (1-p)^R\bigg\{-6+\log(1-p)\bigg(2(R-3)+\log(1-p)\big\{2R-3+(R-1)\log(1-p)\big\}\bigg)\bigg\} \\
-9 (p-1) \big[6+R \log (1-p) \big\{R \log (1-p)+4\big\}\big]\bigg]\bigg)
\label{3d_hetero}
\end{multline}

{\bf Simulations:} We performed cellular automaton type simulations to validate the results given in Eq. \ref{3d_hetero}. Figure \ref{fig:th_g}f shows that the theoretical predictions (Eq.\ref{3d_hetero}) and the simulation results are in excellent agreement. Although the analytic result for $G_{3D}(p,t)$ is complicated, it can be drastically simplified by noting that $P_{br, 3D}\sim 0$ in Eq. \ref{f_expr}. Therefore, the average heterogeneity ($\alpha=1$) in 3D becomes $2x_t(1-x_t)$, which means that the branches (deemed to be linear) evolve independently. The maximum heterogeneity in this case is $\approx 0.5$. Hence, we have a simple expression for $G_{3D}(t,p)$, which can be written as,
\begin{equation}
    G_{3D}(p,t)\approx G_{3D}(x_t)=2x_t(1-x_t).
    \label{simple_g}
\end{equation}
 Since, $P_{br}$ is specific to the geometry of the tissue, we explored the dependence of heterogeneity in different geometries (see Appendix for details).


{\bf Derivation of $F(\alpha)$:} Although Eqs. \ref{3d_hetero} and \ref{simple_g} could be obtained analytically, determination of $\alpha-$dependent $F(\alpha)$ (Eq. \ref{ansatz_ith}) requires simulations.  The term, $F(\alpha)$ in Eq.  \ref{ansatz_ith}, accounts for the effect of changing birth probability, $\alpha$. Figure \ref{fig:theory}a shows ITH as a function of $x_t$ for different $\alpha$ values. The extent of ITH decreases as $\alpha$ decreases. Upon decreasing $\alpha$, the probability of a vacant site in the neighborhood of a cancer cell increases. As a result, the cancer cell gives birth to a daughter cell with similar genetic information. Therefore, the similarity in the genetic information between the neighboring cells increases, which decreases the overall heterogeneity of the tumor.

To extract the functional form of $ F(\alpha)$, we fit $\frac{ITH(x_t=0.5)}{G_{3D}(x_t=0.5)}$ versus $\alpha$ to a line, as shown in Figure \ref{fig:theory}b, which yields, 
\begin{equation}
F(\alpha)=(0.2\alpha+0.77).
\label{eq:hetero2}
\end{equation} 
Since, we obtained Eq. \ref{eq:hetero2} by fitting $\frac{ITH(x_t=0.5)}{G_{3D}(x_t=0.5)}$ at $x_t=0.5$, there is no guarantee that the linear dependence should hold for all values of $x_t$. In order to show that $F(\alpha)$ is accurately given by Eq. \ref{eq:hetero2} for all values of $x_t$, we plotted $\frac{ITH}{F(\alpha)}$ in Figure \ref{fig:theory}c. Figure \ref{fig:theory}c shows that all the curves, corresponding to different values of $\alpha$, collapse onto the same master curve, which coreesponds to $G_{3D}(p, t)$. Note, that $F(\alpha)\approx 1$ when $\alpha=1$. Therefore, the general closed expression for ITH, is the result announced in Eq. \ref{ansatz_ith}. Expression for intra-tumor heterogeneity in Eq.\ref{ansatz_ith} for an exogenous cancer is expressible solely in terms of the probability of gene mutation ($p$) and cell birth ($\alpha$). It is worth reiterating that the simplicity of the analytic result makes it most useful in analyzing the experimental data. 

\subsection{Comparison of theory to M-Seq data}
We calculated ITH \textcolor{black}{of individual patients with skin \cite{harbst2016}, lung \cite{debruin2014, jamal2017tracking}, esophagus \cite{cao2015} or kidney \cite{gerlinger2014} cancers from M-seq data (described in section \rom{2}.A).} In order to compare theory (Eq. \ref{ansatz_ith}) with M-Seq data, $x_t$ and $\alpha$, need to be evaluated for every patient specific M-seq data. 

{\bf Estimating $\alpha$:} During tumor evolution, the cancer cells usually acquire between ($1-10$) driver mutations\cite{martincorena2017universal}, which \textcolor{black}{bestow fitness advantage to} the deleterious cells.  Assuming each driver mutation increases the birth probability by $0.01$ and the number of driver mutations is $5$, we obtain $\alpha=0.55$. 

{\bf Estimating $x_t$:} The parameter, $x_t$, can be directly calculated from the M-Seq data. Note, that $x_t$ refers to the probability that a gene loci is not mutated at $t$. In order to explain the calculation of $x_t$ for each patient, we use Figure \ref{fig:MSeq_figure}b. All M-Seq readouts satisfy $0 \leq x_t \leq1$. For the data in Figure \ref{fig:MSeq_figure}b, the total number of sites is $(n=83)\times7=581$. From Figure \ref{fig:MSeq_figure}b it follows that the total number of non-mutated sites in R2 is 37, R3 is 40, R4 is 42, R6 is 22, R1 is 40, R5 is 41 and R7 is 35. The sum of these numbers $=257$, gives the total non-mutated sites in Figure \ref{fig:MSeq_figure}b, which unifies $x_t=\frac{257}{581}=0.44$.  Similarly, we calculated the $x_t$ value for other patients. 

Using $\alpha=0.55$ and $x_t$, the ITH value can be evaluated (using Eq.\ref{ansatz_ith}) for all the patients. Figure \ref{theory_expt_ith_compare}, that compares the theoretical predictions and M-seq data, shows ITH for patients with exogenous cancers (i.e skin, lung and esophagus) is well captured by our theory. In contrast, the theory drastically overestimates the ITH value for kidney cancer, which is endogenous. 

To quantify the accuracy of the theory in estimating the M-seq data, we use  the Pearson correlation coefficient ($\rho$). The value of $\rho$ for skin cancer is 0.97 with a $p$ value of $10^{-4}$, and 0.88 for lung cancer with a $p$ value of $0.02$ (Figure \ref{theory_expt_ith_compare}a). For the esophageal cancer, we did not calculate the $\rho$ value because the sample has only two patients. Surprisingly, \textcolor{black}{the theory} captures the heterogeneity for the two esophageal cancers accurately. We also compared the theoretical predictions with experiments for a different dataset, where the patients suffered from Non-Small Lung Cancer \cite{jamal2017tracking}, in Figure \ref{theory_expt_ith_compare}b. The $\rho$ value for this dataset was 0.87, which shows that the theory matches experiments.\par 

The value of $\rho$ for kidney cancer is only 0.51 (Figure \ref{theory_expt_ith_compare}c) with a $p$ value 0.13, which implies that ITH behavior is not accurately predicted by the theory. Since kidney cancer is predominantly endogenous, we expect there might be some other factors that are not reflected in the theory, which uses only $\alpha$ and average mutation probability.

\subsection{Spatial distribution of tumor mutation burden (TMB)}
 \label{Ex_vs_En}
  
To discern the reason that the theory is effective in capturing ITH for exogenous cancers, we investigated the spatial distribution of TMB in distinct tumor regions. Figure \ref{cv_skin_lung_eso_kidney}a and \ref{cv_skin_lung_eso_kidney}b show the $x-y$ and $y-z$ cross-section of the simulated tumor. The colors indicate the TMB \textcolor{black}{(number of genes with non-synonymous mutations) in each cell.} The snapshots from tumor cross-sections show, in no uncertain terms, that TMB varies from cell to cell. To extract the spatial distribution of TMB across tumor regions, we divided the tumor into ten regions (similar to M-Seq data), each comprising of $\approx 5,000$ cells. The spatial variation of TMB is illustrated using the coefficient of variation ($c_v$) of TMB across tumor regions,
\begin{equation}
c_v=\frac{\sigma}{\mu},
\end{equation}
where $\sigma$ is the standard deviation of TMB, and $\mu$ is the mean TMB. A small (large) value of $c_v$ indicates low (high) variability. In Figure \ref{cv_skin_lung_eso_kidney}c, we find that the $c_v$ value in simulations is $0.008$. We adopted a similar method to calculate the $c_v$ values from M-Seq data for the four cancer types.


 Figure \ref{cv_skin_lung_eso_kidney}c shows the $c_v$ values for skin \cite{harbst2016}, lung\cite{jamal2017tracking,debruin2014}, esophageal\cite{cao2015} and kidney cancer \cite{gerlinger2014}. The data shows that skin \cite{harbst2016} has an average $c_v \approx 0.023$, implying that the extent of spatial variation in TMB is not significant. The data for Lung cancer\cite{debruin2014} has an average $c_v$ of $\approx 0.086$ ($\approx 8.6\%$) depicting that spatial distribution of TMB in lung cancer is less uniform than skin cancer. However, Figure \ref{cv_skin_lung_eso_kidney}c also shows that the $c_v$ value for lung cancer is large because of patient L002 ($c_v\approx0.274$), which might be an outlier among the six patients. The second Lung cancer dataset\cite{jamal2017tracking} has an average $c_v$ of $\approx 0.066$ ($ 6.6\%$). Skin and lung cancers have approximately uniform spatial distribution of TMB, which resembles the simulation results. Esophageal cancer, which belongs to cancers predominantly caused by exogenous mutagens like skin and lung cancers, has the highest $c_v$ value ($c_v\approx 0.2$). However, there are only two patient data, which prevents us from drawing general conclusions. Surprisingly, kidney cancer dataset\cite{gerlinger2014} also has a very high $c_v$ value of $\approx 0.18$, which is approximately an order of magnitude higher than the $c_v$ for the skin-cancer dataset. \textcolor{black}{ The high TMB variation in endogenous cancers is not captured by the theory, which explains the low $\rho$ value for endogenous cancer. Finding ways to incorporate the heterogeneity in endogenous cancer, and connect it to TMB would be an interesting avenue for future research\cite{li2021imprints}.}
 
 \subsection{Spatial Variations in ITH}
 \label{spat_het}
In the previous sections, we dealt with the average values of intra-tumor heterogeneity. However, a lot of information is lost when considering only average measures of ITH. Most importantly, spatial variations in heterogeneity are smeared out. To illustrate the extent of spatial variations, we consider semi-infinite lattice, infinite lattice and 3D lattice geometry with the birth probability, $\alpha=1$. For the case of semi-infinite case, where growth occurs unidirectionally away from the origin, we begin with a cell at the origin at time $t=0$. Since $\alpha=1$, at every time step, only the cell at the boundary will divide. Therefore, the number of cells, $M(t)=t+1$. We evaluate the average heterogeneity as a function of both inter-cellular distance ($r$) and time ($t$) ($G(r,p,t)$). 

For the case of semi-infinite 1D lattice, $G_{1D,1}(r,p,t)$ is given by 
\begin{equation}
G_{1D,1}(r,p,t)=\frac{1}{M(t)-r}\sum_{i=1}^{M(t)-r}2x_t(1-x_{t-i})
\label{space1}
\end{equation} 
where $M(t)-r$ denotes the number of pairs of cells with inter-cellular distance $r$. In Eq. \ref{space1},the $2x_t(1-x_{t-i})$, is the effective average HD (see Eq. \ref{nonindependent3} in the Appendix) for a pair of cells at $t$ and $(t-i)$. Upon performing the summation, Eq. \ref{space1} becomes,
\begin{equation}
G_{1D,1}(r,p,t)=2x_t\bigg[1-\frac{x_t}{t-r}\bigg\{\frac{1}{p}\bigg(\frac{x_r}{x_t}-1\bigg)\bigg\}\bigg],
\label{space2}
\end{equation} 
where $x_t=(1-p)^t$ and $x_r=(1-p)^r$. Figure \ref{spatial_g}a shows that agreement between simulations and Eq. \ref{space2} is excellent. We observe that ITH is small for smaller inter-cellular distance because majority of cells that are just born are close to their parent, and hence, the daughter cells retain the character of the parent cells. We consider that since small distance incorporate recent birth events they have less heterogeneity compared to cells with larger inter-cellular distance. 

For the infinite 1D lattice, where growth can occur on both the sides of the origin, $G_{1D,2}(r,p,t)$ is given as, 
\begin{equation}
G_{1D,2}(r,p,t)= 
\begin{cases}
    2x_t\bigg[1-\frac{x_t}{2t-r}\bigg\{\frac{2}{p}\bigg(\frac{x_r}{x_t}-1\bigg)+ r\bigg\}\bigg]
,& \text{if } r< t\\
    2x_t(1-x_t), & \text{if } r\geq t
\end{cases}
\label{space3}
\end{equation}
The average spatial and temporal variation of $G_{1D,2}(r,p,t)$ ($\alpha=1$) in Eq. \ref{space3} behaves differently depending on the condition $r<t$ and $r\geq t$. This is because in 1D infinite lattice there are 2 branches. Heterogeneity among cells with inter-cellular distance $r\geq t$, implies that the 2 cells belong to different branches. Therefore, they evolve independently. If $r<t$, we have to incorporate cells within a single branch as well as cells from the 2 different branches. Figure \ref{spatial_g}b shows a good agreement between simulations and theory.

For the 3D lattice, it is difficult to get a closed form expression for the spatial dependence of heterogeneity because the number of branches $\propto 4\pi R^2$, where $R$ is the radius of the tumor. Therefore, we just show simulation results.  Figure \ref{spatial_g}c shows the dependence of heterogeneity as a function of inter-cellular distance. Even in this case, we observe that at large distances the heterogeneity saturates approximately to the value for the independent case. 

{\bf Sub-Sample to Sub-sample variations in a single tumor:} We resorted to 3D cellular automaton simulations to reveal the spatial variations in ITH using $\alpha=0.55$ and $p=0.003$. Figure \ref{hamming_mat} illustrates the massive sample to sample Hamming distance variations in the tumor. Figure \ref{hamming_mat}a shows the Hamming distance map between 2,000 cells, represented as a $2,000\times2,000$ matrix. The matrix shows that different cell pairs have different HD values, representing different levels of  heterogeneity. The zoomed in matrix of size $100\times 100$ in Figure \ref{hamming_mat}b, shows the heterogeneity among cell pairs at a finer resolution. It is clear that the HD values change dramatically varying from $0$ to $0.5$. The HD distance in Figure \ref{hamming_mat}a gradually changes as cells from different regions are sampled, as can seen from Figure \ref{hamming_mat}c (zoomed in Figure \ref{hamming_mat}d) and \ref{hamming_mat}e (zoomed in Figure \ref{hamming_mat}f). Hamming distance matrices in Figures \ref{hamming_mat}c and \ref{hamming_mat}e, are calculated for cells which are approximately $20$ and $27$ lattice units away from the center of the tumor. The color pattern, quantifying the magnitude of Hamming Distance, gradually changes from yellowish to green (Figures \ref{hamming_mat}a, \ref{hamming_mat}c and \ref{hamming_mat}e). Taken together, Figure \ref{hamming_mat} shows the rich heterogeneous spatial patterns that emerges depending on where we probe the tumor. The ITH, provides insights into the complexity of the tumor evolution. The results in Figure \ref{hamming_mat} suggest that if the number of biopsied regions increases, it would reveal far greater changes in $HD$ than is portrayed in Figure \ref{fig:MSeq_figure}c. Of course, there are serious practical limitations when data from human patients is sought.

Figure \ref{pic_het} shows the colorful pattern obtained when the cells, located $\sim 20$ lattice units away from the tumor centre, are painted according to their $HD$ values with respect to the cell at the tumor center. Though tantalizingly beautiful, it shows that quantifying intra-tumor heterogeneity, represented as a HD map, is a challenging problem. 
 


\section{Discussion \& Conclusion}

We developed a statistical mechanical theory, supplemented by simulations, for spatio-temporal variations in ITH associated with cancers. The resulting theoretical expressions are used to analyze M-Seq data on four cancer types obtained from biopsies from multiple regions in a single solid tumor. The agreement between theory and experiments is good for exogenous cancers (skin and lung) but not so for kidney cancers, which is endogenous. The most likely explanation is that factors besides cell division probability ($\alpha$) and mutation probability ($p$), which are the only ingredients in the theory, are relevant. At present, it is unclear how unknown factors could be taken into account. 

\textcolor{black}{Our theory, based on the premise that neutral evolution is valid \cite{tarabichi2018neutral, cannataro2018neutral, mcdonald2018currently, tung2021signatures,wang2018evolution}, uses Hamming Distance ($HD$) as a measure to quantitatively describe genetic variations in M-Seq data.} We show that ITH can be factored into a product of two terms, $G(p, t)$ and $F(\alpha)$, where $G(p,t)$ is the ITH for $\alpha=1$, and $F(\alpha)$ is the scale factor that takes into account that $\alpha$ ($0.5<\alpha<1$) is less than unity. We discovered that $G(p, t)$ depends on the geometry in which the tissue is embedded. Surprisingly, we find in three-dimensions, \textcolor{black}{$G(p, t)=2x_t(1-x_t)$} which coincides with the result for cells that evolve independently. The scale factor $F(\alpha)$ depends linearly on $\alpha$, $F(\alpha)=(0.2\alpha+0.77)$. It is remarkable that the final expression of ITH (Eq.\ref{ansatz_ith}), whose validity is confirmed by simulations, accurately explains the results from M-Seq data obtained from patient data. The theory accurately quantifies ITH in exogenous cancers. The theory predicts that endogenous cancer has one order of magnitude higher $c_v$ value compared to exogenous case which explains the lack of success in explaining ITH in edogenous cancers. 

The cellular and automaton type simulations, which were carried out to confirm the theoretical predictions, vividly illustrate the pervasive nature of ITH. On all length scales in the \textit{in silico} tumor, we find that there are substantial ITH variations, as vividly illustrated in Figure \ref{hamming_mat}. The patterns in the HD map suggests that there are dramatic changes over the same size of tumor regions, as illustrated in Figures \ref{hamming_mat}b, \ref{hamming_mat}d, and \ref{hamming_mat}f. We find it remarkable that the 2D matrices associated with the HD map reveal striking dissimilarity, as is evident from Figure \ref{pic_het} and additional plots in the Figure S13. These figures show visually that the ITH in distinct regions are vastly different, suggesting the mere coarse-grained representation available from experiments (Figure \ref{fig:MSeq_figure}b). We should caution the reader that the large scale heterogeneity that is visualized in the simulations may not be revealed (or even present) in solid tumors because it would ultimately require sequencing at the single cell resolution \cite{navin2011tumour}. In addition to spatial heterogeneity distribution, the heterogeneity distribution also evolves in time as is reflected in Figure S14. 


\textcolor{black}{In addition to the genetic heterogeneity discussed above, phenotypic variations \cite{meacham2013tumour,li2021mathematical} of cancer cells are also important,  which is crucial in devising modern personalized cancer therapeutics \cite{moscow2018evidence}.  Recent advancements in imaging modalities have helped unearth heterogeneous physical characteristics at the single cell resolution in three-dimensional cell collectives \cite{valencia2015collective, han2020cell, martino2019wavelength}. Our previous studies have shown that cell-division and apoptosis induced self-generated forces give rise to phenotypic diversity within a tumor \cite{malmi2018cell, sinha2020self,sinha2020spatially,samanta2020far, sinha2021inter, malmi2021adhesion, sinha2021memory}. In principle, the current theoretical framework, could be extended to estimate the ITH for both genetic and phenotypic heterogeneity of evolving tumors.}

{\bf Acknowledgements}
This work is supported by a grant from the National Science Foundation (PHY 17-08128 and PHY-1522550). Additional support was provided by the Collie-Welch Reagents Chair (F-0019).

\section{Appendix}
Motivated by the data in Figures S2-S6 in SI, we developed a statistical mechanical (or probabilistic) theory for quantifying ITH, which is described in the main text. Here, we provide the technical details.  

{\bf Heterogeneity in independently evolving cells:} Consider two cells, $1$ and $2$, which evolve independently (Figure S7a in SI). The elements in the string $X^1$ and $X^2$, which are the genes,  are initialized to zero at $t=0$. This implies that there are no mutations at $t = 0$. Given the rules for acquiring mutations (described in the Methods section in the main text), we evaluate the Hamming distance ($\langle HD[X^1(t),X^2(t)] \rangle$), where $\langle ... \rangle$ represents the ensemble average. We generate an ensemble of DNA strings with mutations from an evolutionary trajectory over a certain time. At each time step, a DNA string associated with a cell, can acquire mutations. The set of mutations acquired at the end of the evolutionary period constitutes the ensemble.

The Hamming distance ($\langle HD[X^1(t),X^2(t)] \rangle$) is,

\begin{equation}
\langle HD[X^1(t),X^2(t)]  \rangle = \sum_{k=1}^{n}\frac{k}{n}\{\mathbb{P}[HD[X^1(t),X^2(t)] =\frac{k}{n}]\}
\label{heteroindependent}
\end{equation} 
 $\mathbb{P}[HD[X^1(t),X^2(t)] =\frac{k}{n}]$ is the probability that the HD between cell $1$ and $2$ is $\frac{k}{n}$ at $t$. We divide by the string length, $n$, because the sites on the string are independent of each other. Moreover, in the M-seq experiments the string length $n$, varies from patient to patient. Therefore, dividing by $n$ enables us to treat each patient data on the same footing. 
 
 The expression for the right hand side of Eq. \ref{heteroindependent} is,
\begin{equation}
\begin{split}
\mathbb{P}[HD[X^1(t),X^2(t)] =\frac{k}{n}] & \\
&=\sum_{\substack{z_0+z_1+z_2+z_{12}=n \\ z_1 + z_2 = k}} \frac{n\, !}{z_0\,! z_1\,! z_2\,! z_{12}\,!} (1-p)^{t(2z_0+z_1+z_2)}[1-(1-p)^{t}]^{2z_{12}+z_1+z_2},
\end{split}
\label{combinatorial}
\end{equation}
where $z_0$ is the number of common sites in $X^1$ and $X^2$ that are not mutated, and $z_{12}$ denotes all the common sites in $X^1$ and $X^2$ that are mutated. The number of sites where $X^1$ is mutated but the corresponding site in $X^2$ is not mutated is $z_1$. Similarly, $z_2$ denotes the number of sites where $X^2$ is mutated but the corresponding site in $X^1$ is not mutated. We also have the constraints that $z_1+z_2=k$, and $z_0+z_1+z_2+z_{12}=n$. In Eq. \ref{combinatorial}, the term $(1-p)^{t(2z_0)}$ accounts for the probability that $z_0$ common sites are not mutated, which is the product $(1-p)^{t(z_0)}(1-p)^{t(z_0)}$. Similarly, in Eq. \ref{combinatorial}, $(1-(1-p)^t)^{2z_{12}}$ is the probability that $z_{12}$ common sites are mutated $[1-(1-p)^t]^{z_{12}}[1-(1-p)^t]^{z_{12}}=[1-(1-p)^t]^{2z_{12}}$.

We can calculate the right hand side of Eq. \ref{combinatorial} to get,
\begin{equation}
\begin{split}
\mathbb{P}[HD[X^1(t),X^2(t)] =\frac{k}{n}]=\binom{n}{k}\{2x_{t}(1-x_{t})\}^k\{x_{t}^2+(1-x_{t})^2\}^{n-k},
\end{split}
\label{probindependent}
\end{equation}
where $x_{t}=(1-p)^{t}$ is the probability that a site has not been mutated till $t$. We substitute Eq. \ref{probindependent} in the R.H.S of Eq. \ref{heteroindependent} and obtain,
\begin{equation}
\begin{split}
\langle HD[X^1(t),X^2(t)]  \rangle=\sum_{k=1}^{n}\frac{k}{n}\binom{n}{k}\{2x_{t}(1-x_{t})\}^k\{x_{t}^2+(1-x_{t})^2\}^{n-k}.
\end{split}
\label{heteroindependent2}
\end{equation}
Surprisingly, equation \ref{heteroindependent2} reduces to a very simple form given by, 
\begin{equation}
\langle HD[X^1(t),X^2(t)]  \rangle=2x_{t}(1-x_{t}).
\label{heteroindependent3}
\end{equation}
Eq. \ref{heteroindependent3} is the average heterogeneity among independently evolving cells. In our case, this type of evolution refers to cells on two different 1D semi-infinite branches. Figure S7b in the SI shows that the theoretical prediction given in Eq.\ref{heteroindependent3} is in excellent agreement with the simulation results.

{\bf Heterogeneity for non-independent evolving cells:} We now consider a scenario when two cells have a common ancestor at time, $t>0$. Let us consider the case when cell $1$ evolves in time, acquiring exogenous mutations, from $t=0$ till $t$. After time $t$, it divides and gives birth to cell $2$. During the birth process, the information contained in string $X^1$ is copied to string $X^2$. Figure S8 in the SI illustrates the correlated evolutionary dynamics. 
 
 Our goal is to evaluate the average heterogeneity ($\langle HD[X^1(t+s),X^2(t+s)] \rangle$) between cells $1$ and $2$ at a later time $=t+s$.  As before, $\langle ... \rangle$ represents the ensemble average. Following the same method used to derive Eq. \ref{heteroindependent3}, we obtain,
\begin{equation}
\begin{split}
\mathbb{P}[HD[X^1(t+s),X^2(t+s)] =\frac{m}{n}]=\binom{l}{m}\{2x_{s}(1-x_{s})\}^m\{x_{s}^2+(1-x_{s})^2\}^{l-m},
\end{split}
\label{nonindependent}
\end{equation}
where $x_s=(1-p)^s$, and $l$ is the number of non-mutated sites in string $X^1(t)$. It can be shown that $\langle l \rangle=n(1-p)^t$. Using a similar expression, as in Eq. \ref{heteroindependent}, we can write, 
\begin{equation}
\langle HD[X^1(t+s),X^2(t+s)]  \rangle = \sum_{m=1}^{l}\frac{m}{n}\{\mathbb{P}[HD[X^1(t),X^2(t)] =\frac{m}{n}]\}.
\label{nonindependent2}
\end{equation} 
By substituting Eq. \ref{nonindependent} in the R.H.S of Eq. \ref{nonindependent2}, we obtain,
\begin{equation}
\langle HD[X^1(t+s),X^2(t+s)]  \rangle=2x_{s}(1-x_{s})x_t.
\label{nonindependent3}
\end{equation}
The above equation can also be written as $\langle HD[X^1(t+s),X^2(t+s)]  \rangle=2x_{t+s}(1-x_{s})$. Equation \ref{nonindependent3}, represents the average heterogeneity between two cells which had common evolution till time $t$ but evolved independently for the subsequent time interval, $s$. It can be shown from the expression given above that if we set $t=0$, we get back the equation for independent evolution, derived in Eq.\ref{heteroindependent3}. The result in Eq. \ref{nonindependent3} is very interesting because it shows that the temporal correlations between the two cells arises because of the copying mechanism during cell-division, which also implies that the heterogeneity is smaller compared to the independently evolving cells (Eq.\ref{heteroindependent3}).

{\bf Branching increases intra-tumor heterogeneity:}
The heterogeneity measure between cells that evolve independently and in a correlated manner, allows us to compute the average heterogeneity withing an evolving tumor for $\alpha=1$, which we denote as $G(p,t)$. We present the results for $G(p,t)$ for various physical geometries: $G_{1D,1}(p,t)$ for semi-infinite lattice , $G_{1D,2}(p,t)$ for infinite lattice, $G_{2D}(p,t)$ for 2D lattice, and $G_{3D}(p,t)$ for 3D lattice with birth probability $\alpha=1$. We consider these geometries because the number of branches in an evolutionary tree depends on its geometrical structure. The M-Seq data shows that the phylogeny tree of a tumor consists of many branches \cite{gerlinger2012intratumor}, in contrast to the linear evolution model suggested by Nowell \cite{nowell1976clonal}. Thus, by considering different physical geometries representing the branches, we can probe the role of branching on ITH. 

To calculate $G(p,t)$, we first introduce $F(r_1,r_2,t)$, which is the average heterogeneity for pair of cells at distance $r_1$ and $r_2$ from the origin at time $t$. Without loss of generality, we assume that $r_1 \geq r_2$. The general expression for $F(r_1,r_2,t)$ for any geometry is given by,
\begin{equation}
F(r_1,r_2,t)=P_{br}\{2x_t(1-x_{t-r_2})\}+(1-P_{br})\{2x_t(1-x_t)\},
\label{heteroaverage1}
\end{equation}
where $P_{br}$ is the probability that both the cells belong to the same branch of the evolutionary tree. For the 1D semi-infinite lattice, $P_{br}=1$ because there is only one branch. For the 1D infinite lattice, $P_{br}=\frac{1}{2}$. For a 2D lattice, $P_{br}=\frac{(\pi r_1^2)r_1}{(\pi r_1^2)(\pi r_1^2)}=\frac{1}{\pi r_1}$, and for the 3D lattice, $P_{br}=\frac{(\frac{4}{3}\pi r_1^3)r_1}{(\frac{4}{3}\pi r_1^3)(\frac{4}{3}\pi r_1^3)}=\frac{1}{\frac{4}{3}\pi r_1^2}$.  In 2D and 3D, the number of cells within a radius $r_1$ is proportional to $\pi r_1^2$ and $\frac{4}{3} \pi r_1^3$ respectively, and the number of cells that lie on the same branch is proportional to $r_1$ as long as the branches are linear. Therefore, we obtain the above expressions for $P_{br}$ in 2D and 3D. Note that to arrive at the expression for $P_{br}$, we have assumed that the evolutionary tree is comprised of many semi-infinite lattices with zero curvature. It is important to note that for 2D and 3D case, $P_{br}\rightarrow 0$ for $r_1>>1$. Interestingly, $F(r_1, r_2, t)$ does not depend on $r_1$ but only on $r_2$ for 1D semi-infinite or infinite lattice.
Eq. \ref{heteroaverage1} has two parts, which we derived in Eqs. \ref{heteroindependent3} and \ref{nonindependent3}. The first part corresponds to cells that undergo correlated evolution while the second part refers to cells undergoing independent evolution. Therefore, in all the possible geometries, $F(r_1,r_2,t)$ takes the following form, 
\begin{equation}
F(r_1,r_2,t)= 
\begin{cases}
    2x_t(1-x_{t-r_2}), & \text{ 1D semi-infinite} \\
    \frac{1}{2}2x_t(1-x_{t-r_2})+(1-\frac{1}{2})2x_t(1-x_t), & \text{ 1D infinite}\\
    \frac{1}{\pi r_1}2x_t(1-x_{t-r_2})+(1-\frac{1}{\pi r_1})2x_t(1-x_t), & \text{ 2D}\\
      \frac{1}{\frac{4}{3}\pi r_1^2}2x_t(1-x_{t-r_2})+(1-\frac{1}{\frac{4}{3}\pi r_1^2})2x_t(1-x_t), & \text{ 3D}
\end{cases}
\label{heteroaverage2}
\end{equation}
To compute $G(p,t)$ (Eq.(2) in the main text), we need to integrate $F(r_1,r_2,t)$ over space with appropriate normalization. The expression for $G(p,t)$ in different geometries are given by,
\begin{equation}
G(p,t)=
\begin{cases}
G_{1D,1(2)}(p,t)=\frac{\int _1^R\int _1^{r_1}F(r_1,r_2,t)dr_2dr_1}{\int _1^R\int _1^{r_1}dr_2dr_1}, & \text{if 1D semi-infinite or infinite lattice.}\\
G_{2D}(p,t)=\frac{\int _1^R\int _1^{r_1}r_1r_2 F(r_1,r_2,t)dr_2dr_1}{\int _1^R\int _1^{r_1}r_1r_2 dr_2dr_1}, & \text{if 2D lattice.}\\
G_{3D}(p,t)=\frac{\int _1^R\int _1^{r_1}r_1^2r_2^2 F(r_1,r_2,t)dr_2dr_1}{\int _1^R\int _1^{r_1}r_1^2r_2^2dr_2dr_1} , & \text{if 3D lattice.}
\end{cases}
\label{heteroaverage3}
\end{equation}
In the above equation, $R$ is the tumor radius in all the geometries except semi-infinite lattice (for semi-infinite case $R$ is the length of the tumor). On integrating the above equations in Eq. \ref{heteroaverage3} using Mathematica, $G(p,t)$ in all the geometries is given by,
\begin{equation}
G_{1D,1}(p,t)= \frac{2(1-p)^t}{(R-1)^2} \bigg((R-1)^2-\frac{(1-p)^{t-R-1}}{\log^2(1-p)} \bigg\{2 (1-p)^R \bigg[(R-1) \log (1-p)-1\bigg]-2(p-1)\bigg\}\bigg)
\label{heteroaverage4}
\end{equation}

\begin{multline}
G_{1D,2}(p,t)= \\
\frac{2(1-p)^t}{(R-1)^2} \bigg(\frac{(1-p)^{t-R-1}}{\log ^2(1-p)} \bigg\{(1-p)^R+(p-1)-(R-1) (1-p)^R \log (1-p)\bigg\}-\frac{1}{2}(R-1)^2 \bigg[(1-p)^t-2\bigg]\bigg)
\label{heteroaverage5}
\end{multline}

\begin{multline}
G_{2D}(p,t)= \\
\frac{8 (1-p)^t}{\pi  \left(R^2-1\right)^2}\bigg\{-\frac{1}{12}(R-1)^2\bigg[3 \pi (R+1)^2\bigg((1-p)^t-1\bigg)-4(R+2)(1-p)^t\bigg] \\
  -\frac{(1-p)^{t-R-1}}{\log ^3(1-p)}\bigg(-2 (p-1) \big[2+R \log
   (1-p)\big]+2(1-p)^R\bigg[-2+\\
  \log(1-p)\bigg\{(R-1)\log(1-p) +(R-2)\bigg\} \bigg]\bigg)\bigg\}
 \label{heteroaverage6}
\end{multline}

\begin{multline}
G_{3D}(p,t)=\\
\frac{3 (1-p)^t}{\pi  \left(R^3-1\right)^2}\bigg(-\frac{2}{3} \pi  \left(R^3-1\right)^2 \big[(1-p)^t-1\big]+\frac{3}{4} (R-1)^2 (1-p)^t \left(R^2+2R+3\right) \\
-\frac{(1-p)^{t-R-1}}{\log ^4(1-p)}\bigg[9 (1-p)^R\bigg\{-6+\log(1-p)\bigg(2(R-3)+\log(1-p)\big\{2R-3+(R-1)\log(1-p)\big\}\bigg)\bigg\} \\
-9 (p-1) \big[6+R \log (1-p) \big\{R \log (1-p)+4\big\}\big]\bigg]\bigg)
\label{heteroaverage7}
\end{multline}

 {\bf Comparison beween Theory and Simulations:} To validate the theoretical predictions, we compared the results in Eqs. \ref{heteroaverage4}, \ref{heteroaverage5} and \ref{heteroaverage7} using simulations. Figure \ref{fig:th_g}d shows excellent agreement between equation (\ref{heteroaverage4}) and simulations for the 1D semi-infinite lattice. The 1D semi-infinite lattice can be thought of as a branch in the 3D evolutionary tree. Therefore, as expected the heterogeneity among cells within the same branch is smaller compared to cells that evolve independently. The maximum average heterogeneity when the cells evolve independently, as can be gleaned from Eq. (\ref{heteroindependent3}), is $0.5$ whereas the maximum average heterogeneity among cells within a branch is $\approx 0.36$. The substantial reduction in heterogeneity occurs because of correlations among the cells arising from copying genetic information from parent to child during cell division.

Similarly, for tumor evolution in 1D infinite lattice, the agreement between theory (Eq. \ref{heteroaverage4}) and simulations is excellent (Figure \ref{fig:th_g}e). The 1D infinite lattice comprises of 2 branches with an angle of \ang{180} between the two. The nodes on the two branches are completely independent of one another. However, due to the non-independent evolution of cells within the two branches, the heterogeneity is small compared to the independent evolution case. The maximum heterogeneity is $\approx 0.43$ (Eq. \ref{heteroaverage4}). 

Evolution of tumor in 3D lattice is the most interesting case, and is most relevant. Figure \ref{fig:th_g}f again shows good agreement between theory (equation \ref{heteroaverage7}) and simulations. It is surprising that in 3D, the heterogeneity is similar to the case of independent evolution. This can be understood with the help of equation \ref{heteroaverage1} with $P_{branch}\rightarrow 0$. In this limit, the average heterogeneity behaves like the case for independent evolution (i.e $2x_t(1-x_t))$. The maximum heterogeneity in this case is $\approx 0.5$.

Having shown that our theoretical results are consistent with simulations, we compare the theoretical predictions for semi-infinite, infinite and 3D lattice together. We can clearly see from Figure S9d that $G_{1D,1}(p,t)<G_{1D,2}(p,t)<G_{3D}(p,t)$. The inequality follows because as we increase the dimensionality, we give way to more branches which undermines the effect of reduction of heterogeneity due to copying of genetic information during cell division. Therefore, $G_{3D}(p,t)$ is approximately similar to the case of independent evolution.

\clearpage

\clearpage
\begin{figure}
\includegraphics[clip,width=1\textwidth]{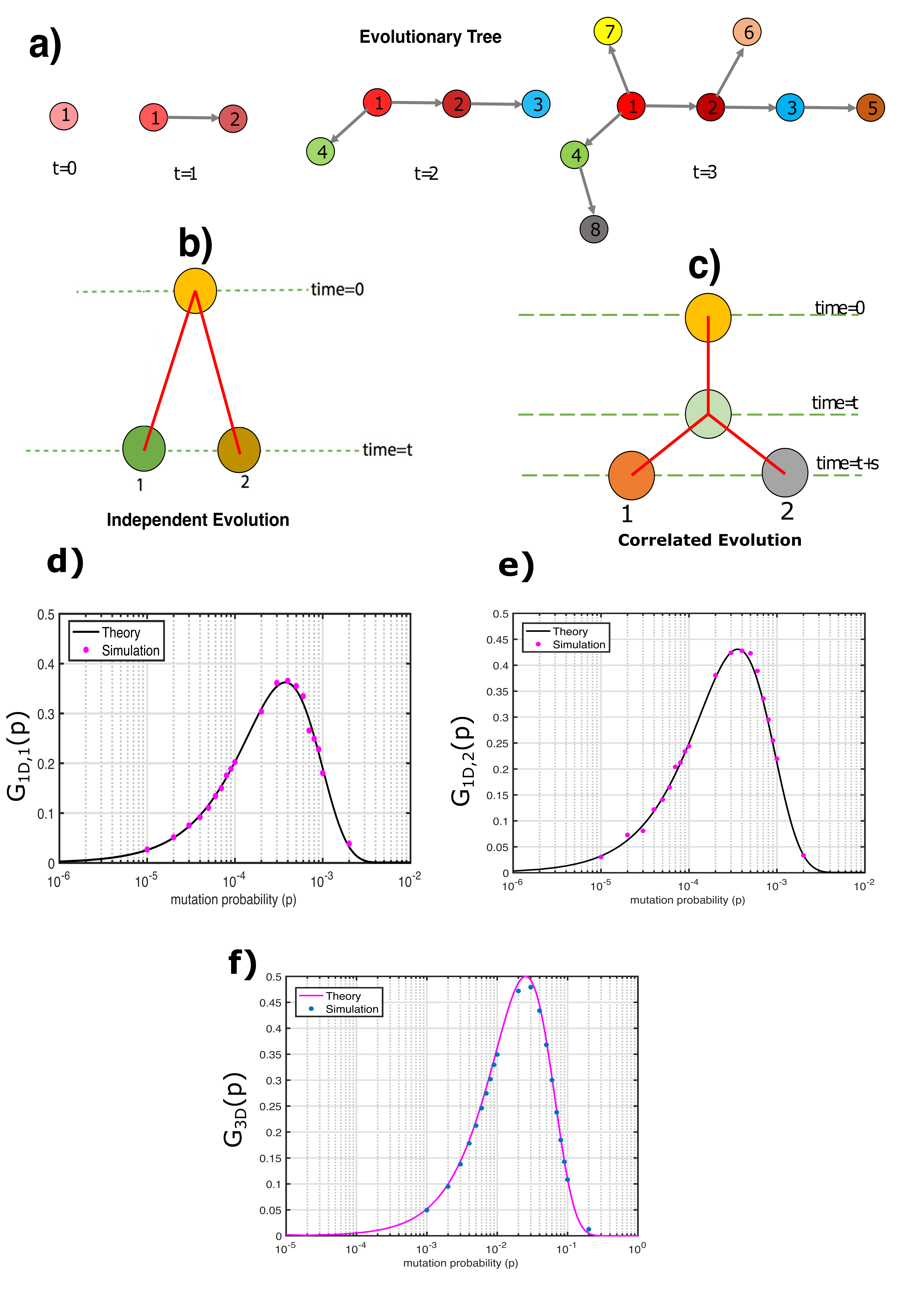}
\caption{(Contined on the following page) }
\label{fig:th_g}
\end{figure}

\clearpage
\begin{figure}
\contcaption{{\bf Calculation of G(p,t) for, $\alpha$, the birth probability set to unity.} {\bf (a)} A schematic of evolutionary tree at time, $t = 0, 1, 2$ and $3$. The cell colors (only for illustration purpose) gradually changes from transparent to opaque as they acquire mutations. The directed edges denote the child-parent relationship between two cells. For instance, the edge from $3\rightarrow 5$ indicates that cell $5$ was born from cell $3$. The evolutionary tree comprises of several branches. A branch is a unique path traversed from the origin (cell $1$) to any leaf node (no directed edge from a node) by following the directed edges. In Figure \ref{fig:th_g}a, $1\rightarrow2\rightarrow3\rightarrow5$ is
the longest branch at t=3. {\bf (b)} Cartoon depicting independent evolution of two cells (green and brown) labeled as $1$ and $2$. The two cells are normal at time $t=0$ and evolve independently acquiring mutations. {\bf (c)} Schematic of correlated evolution of 2 cells (grey and orange). Evolution begins at $t=0$ with a single normal cell (yellow). It evolves till $t$, acquiring mutations (light green) in the process. At $t$, the cell gives birth to a daughter cell. After $t$, the 2 cells evolve independently for the next $s$ time steps. By correlated evolution we mean that the two cells have a common ancestor. In this instance, the common ancestor for the orange and grey cells at $t+s$ is the green cell at $t$. {\bf (d)} Average ITH within an evolving tumor in 1D semi-infinite lattice ($G_{1D, t}(p,t)$). The dots in magenta correspond to simulations with $\alpha=1$ calculated by evolving for $t = 2,000$ time steps. The black line corresponds to Eq. \ref{heteroaverage4} in the SI with $R=t=2,000$. {\bf (e)} \textcolor{black}{Same as (d) but} in 1D infinite lattice ($G_{1D,2}(p,t)$). The black line is a plot of Eq. \ref{heteroaverage5} in the SI with $R=t=2000$. {\bf (f)} \textcolor{black}{Same as (d) but} in 3D lattice ($G_{3D}(p, t)$) for $t = 27$. The systems evolution was carried out for $27$ time steps. The magenta line corresponds to Eq. \ref{3d_hetero} with $R=t=27$.}
\end{figure}

\clearpage
\begin{figure}
\includegraphics[clip,width=1\textwidth]{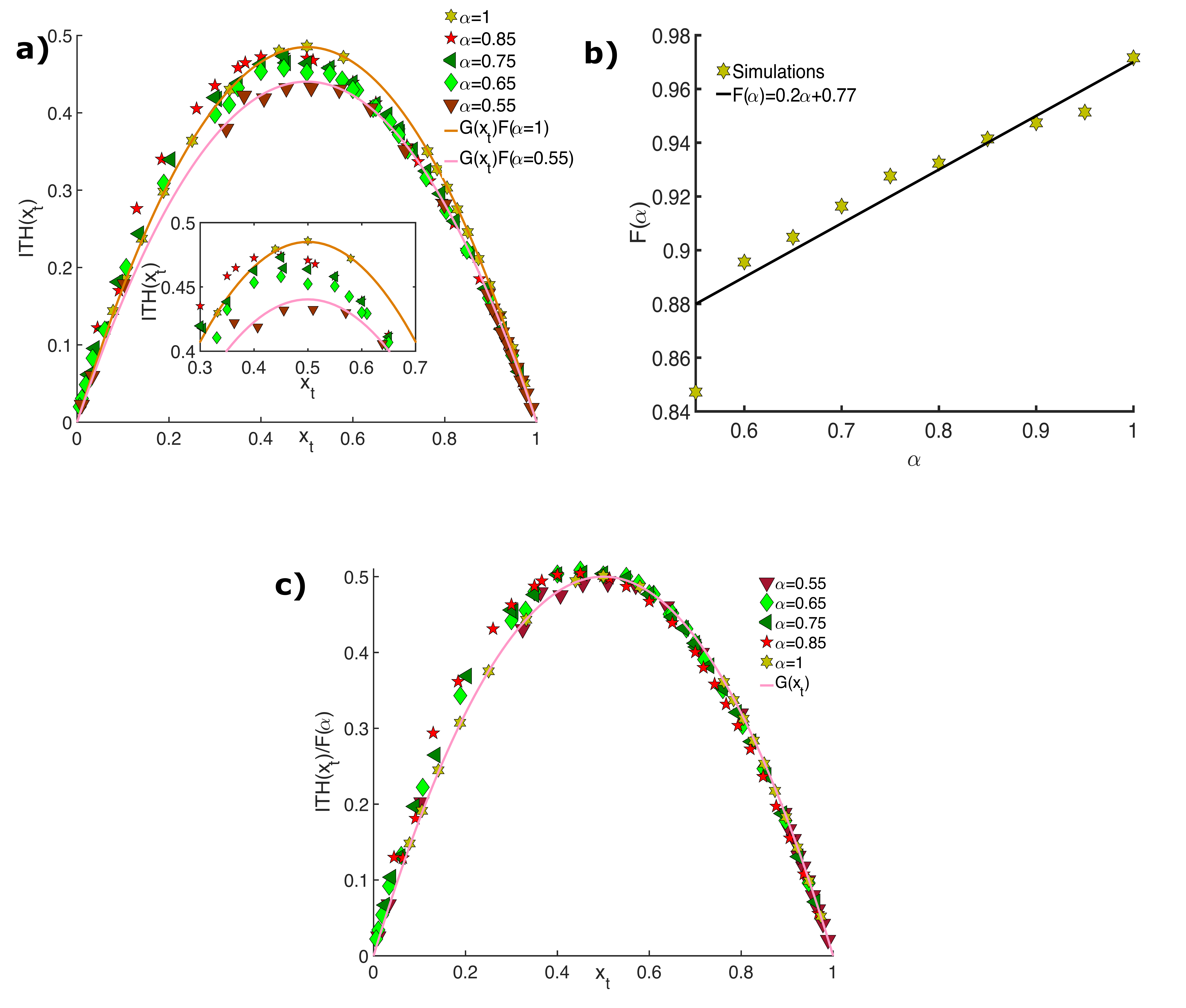}
\caption{{\bf Theoretical Predictions for Intra-tumor Heterogeneity} {\bf (a)} ITH (Eq. \ref{ansatz_ith}) as a function of $x_t$ (probability a gene is not mutated) at $t$. Simulation results are shown for five values of $\alpha=1, 0.85, 0.75, 0.65, 0.55$, the birth probability. Reduction in $\alpha$ decreases the peak value of ITH, as is clear from the inset. {\bf (b)} Plot of $F(\alpha)$ vs $\alpha$. $F(\alpha)=\frac{ITH(x_t=0.5)}{G_{3D}(x_t=0.5)}$, was fit to a line ($F(\alpha)=0.2\alpha+0.77)$), shown in black. {\bf (c)} $\frac{ITH}{F(\alpha)}$, as a function of $x_t$ shows the collapse of the data sets onto a master curve, $G(x_t)=2x_t(1-x_t)$.}
\label{fig:theory}
\end{figure}

\clearpage
\begin{figure}
\includegraphics[clip,width=1\textwidth]{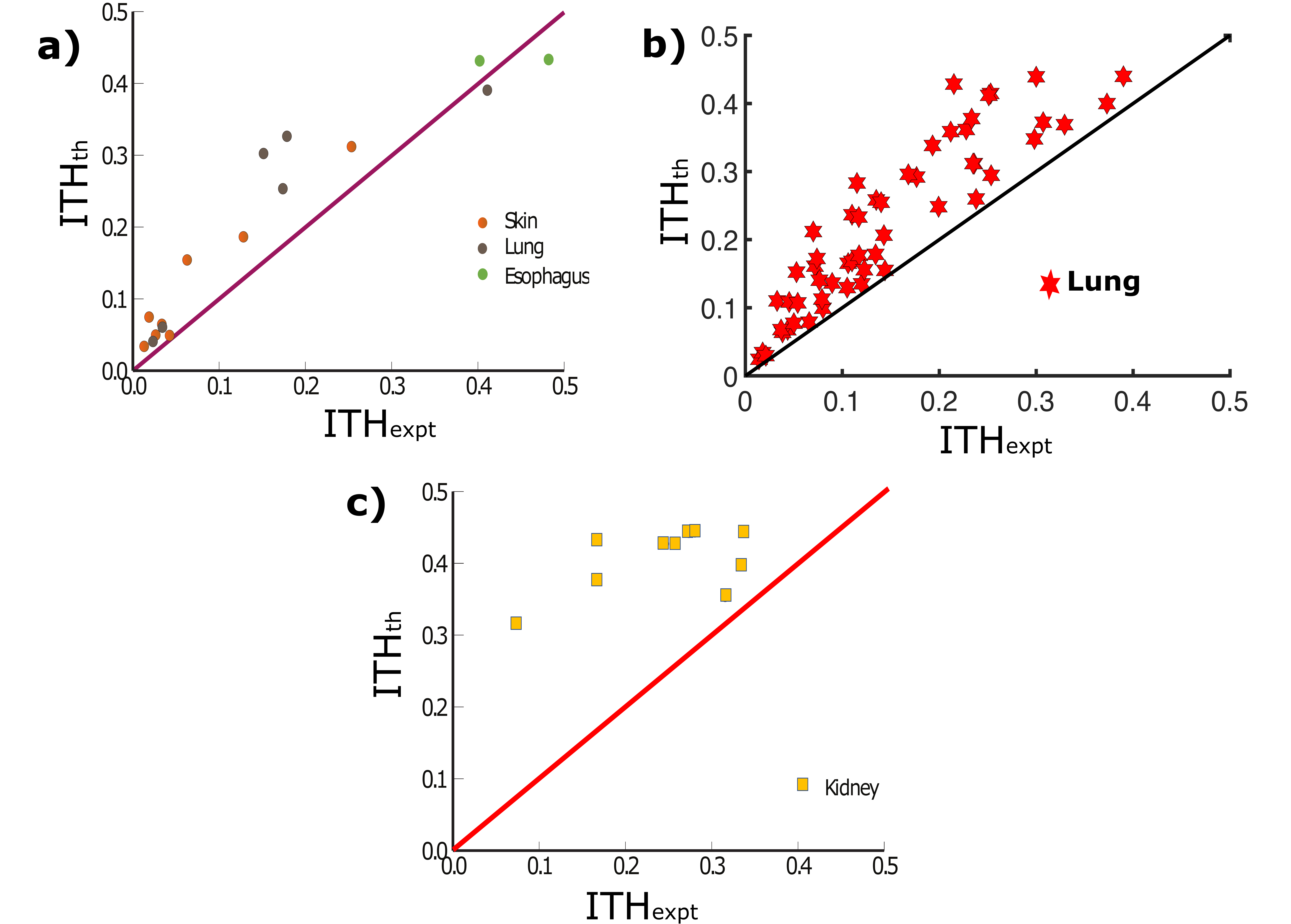}
\caption{{\bf Comparison between theory and experiments (M-seq data)} {\bf (a)} The orange, brown and green colors represent skin, lung, esophageal cancer respectively. The maroon straight line represents perfect linear relation between theory and experiments. The theory captures the heterogeneity of exogenous cancers (skin, lung, esophagus) reasonably accurately. The value of $\rho$ (the Pearson Correlation coefficient) for skin cancer is 0.97 with a $p$ value of $10^{-4}$ and it is 0.88 for lung cancer with a $p$ value of 0.02. {\bf (b)} Same as (a) except the plot is for lung cancer dataset from Jamal et.al\cite{jamal2017tracking}. The value of $\rho$ for this dataset is 0.87. {\bf (c)} ITH for the endogenous kidney cancer, shown in yellow squares, which is an endogenous cancer is not captured well by the theory. For kidney cancer, the value of $\rho$ is 0.5 with a $p$ value 0.13.}
\label{theory_expt_ith_compare}
\end{figure}

\clearpage

\begin{figure}[h]
\includegraphics[clip,width=1\textwidth]{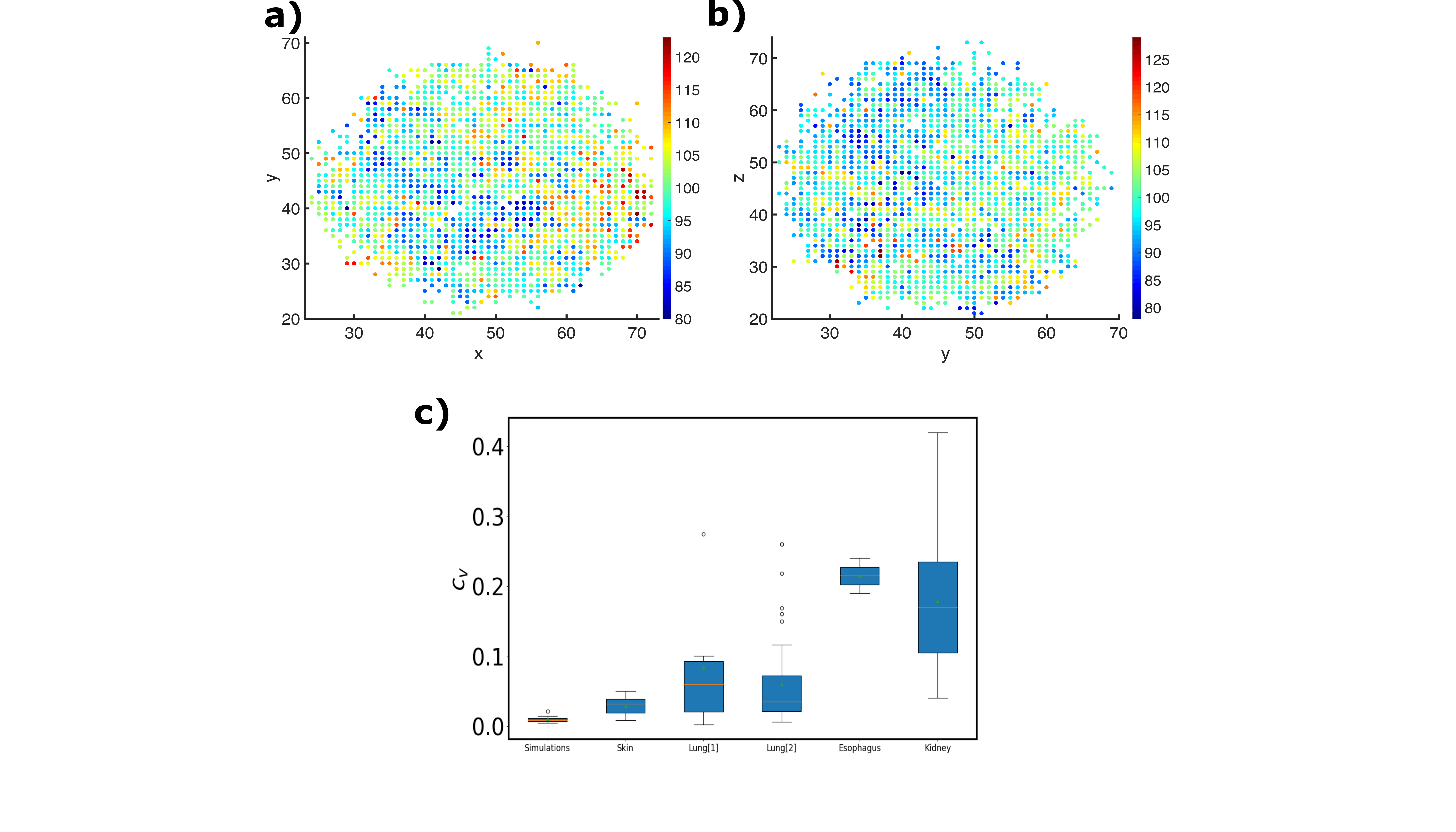}
\caption{{\bf Spatial distribution of mutations} {\bf (a), (b)} $x-y$ and $y-z$ cross-section of the simulated 3D tumor. Distinct colors (see the color bar on the right) indicate the Tumor mutation burden (TMB) on each cell. Large variation in TMB occur throughout the simulated tumor. {\bf (c)} Coefficient of variation, $c_v$, of TMB in simulation and the four cancer types. The $c_v$ $\approx 0.008$, value in the simulations is small. For skin cancer dataset from Harbst et.al \cite{harbst2016}, the $c_v$ value is $\approx 0.023$. For lung cancer dataset from de Bruin et.al. \cite{debruin2014}, the $c_v$ value is $\approx 0.086$. The second Lung cancer dataset from Jamal et.al.\cite{jamal2017tracking} has an average $c_v$ of $\approx 0.066$. Esophageal cancer dataset from Cao et. al\cite{cao2015} has the highest $c_v$ value among the four cancer types ($c_v\approx 0.2$). The kidney cancer dataset from Gerlinger et. al \cite{gerlinger2014} has a high $c_v$ value of $\approx 0.178$.}
\label{cv_skin_lung_eso_kidney}
\end{figure}

\clearpage
\begin{figure}[h]
\includegraphics[width=19.5cm]{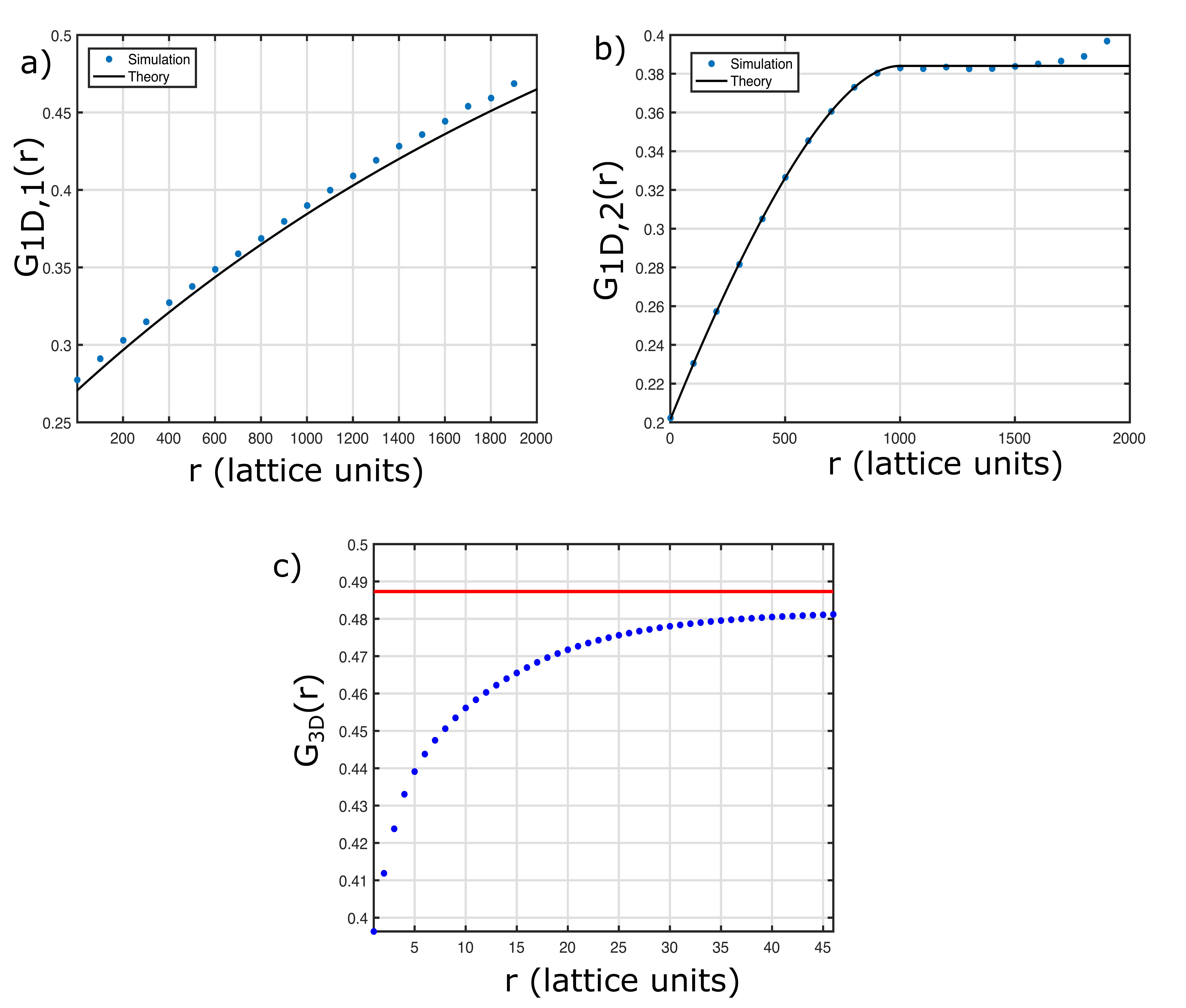}
\caption{{\bf Spatial dependence of G(p) for different geometries}. ({\bf a}) Heterogeneity ,$G_{1D,1}(r)$ (Eq. \ref{space2} evaluated at $p=0.0005$ and $t=2,000$), as a function of inter-cellular distance within a 1D semi-infinite lattice for $\alpha=1$. The black line is the result from theory (Eq.\ref{space2}). The spatial dependence of $G_{1D,1}(r)$ arises because of the copying mechanism during cell-division, which introduces correlations between cells. ({\bf b}) Heterogeneity, $G_{1D,2}(r)$ (Eq. \ref{space3} evaluated at $p=0.0003$ and $t=1,000$)), as a function of inter-cellular distance for a 1D infinite lattice. The blue dots correspond to simulations. (Continued on the next page)}
\label{spatial_g}
\end{figure}

\clearpage
\begin{figure}
\contcaption{The black line is the theoretical result (Eq.\ref{space3}). The curve saturates to the maximum allowed heterogeneity value given by equation $2x_t(1-x_t)$.  ({\bf c}) Heterogeneity, $G_{3D}(r)$  as a function of $r$ within a 3D tumor for $\alpha=1$. The dots in blue correspond to simulation for $p=0.02$. The tumor was evolved for $27$ time steps, and had $\approx 50,000$ cells at the end of the simulations. The diameter of tumor is $\approx 46$ lattice units. The red line corresponds to $2x_t(1-x_t)$ and is the maximum possible ITH value.}
\end{figure}

\clearpage
\begin{figure}[h]
\includegraphics[width=19cm]{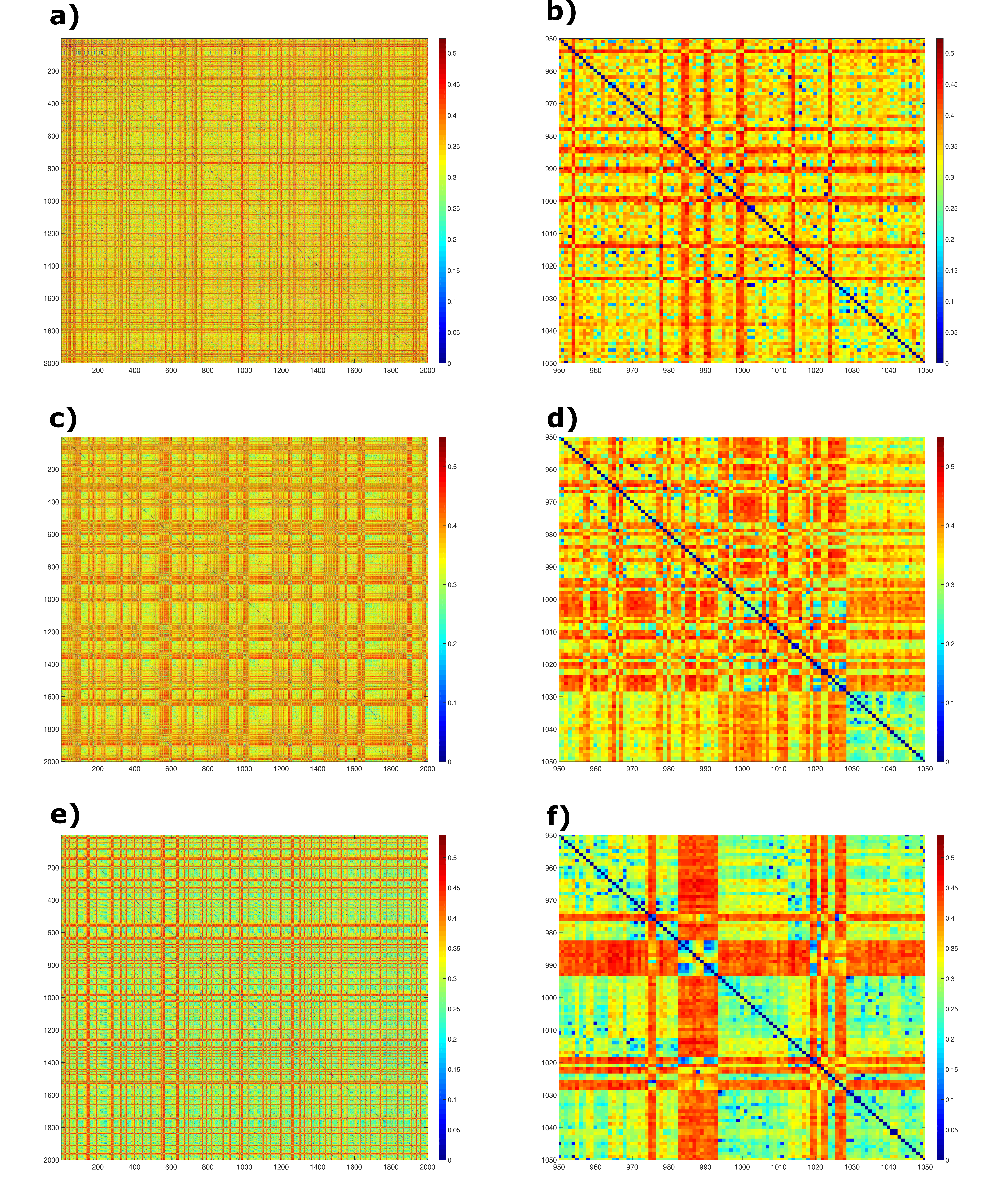}
\caption{{\bf Hamming distance matrices for regions in the simulated tumor for $\alpha=0.55$, $p=0.003$ and $t = 113$}.(Continued on the next page)}
\label{hamming_mat}
\end{figure}

\clearpage
\begin{figure}
\contcaption{{\bf (a)} Hamming matrix for 2,000 cells which are closest to the center of the simulated tumor. Each element ($i,j$) of the matrix gives the values of the HD between a pair of cells. The color bar on the right corresponds to the HD values. {\bf (c)} Same as (a) but for cells which are located approximately 20 lattice units away from the center of the tumor. {\bf (e)} Same as (a) and (c) but for cells which are located approximately 27 lattice units away from the center of the tumor. {\bf (b, d, f)} Zoomed in portion of size $100\times100$ from $950$ to $1050$ in Figures (a, c, e). The figures illustrate the sub-sample to sub-sample variations in the HD values depending on the location of the region sampled. The scales for HD are given on the right.}
\end{figure}

\clearpage
\begin{figure}[h]
\includegraphics[width=15cm]{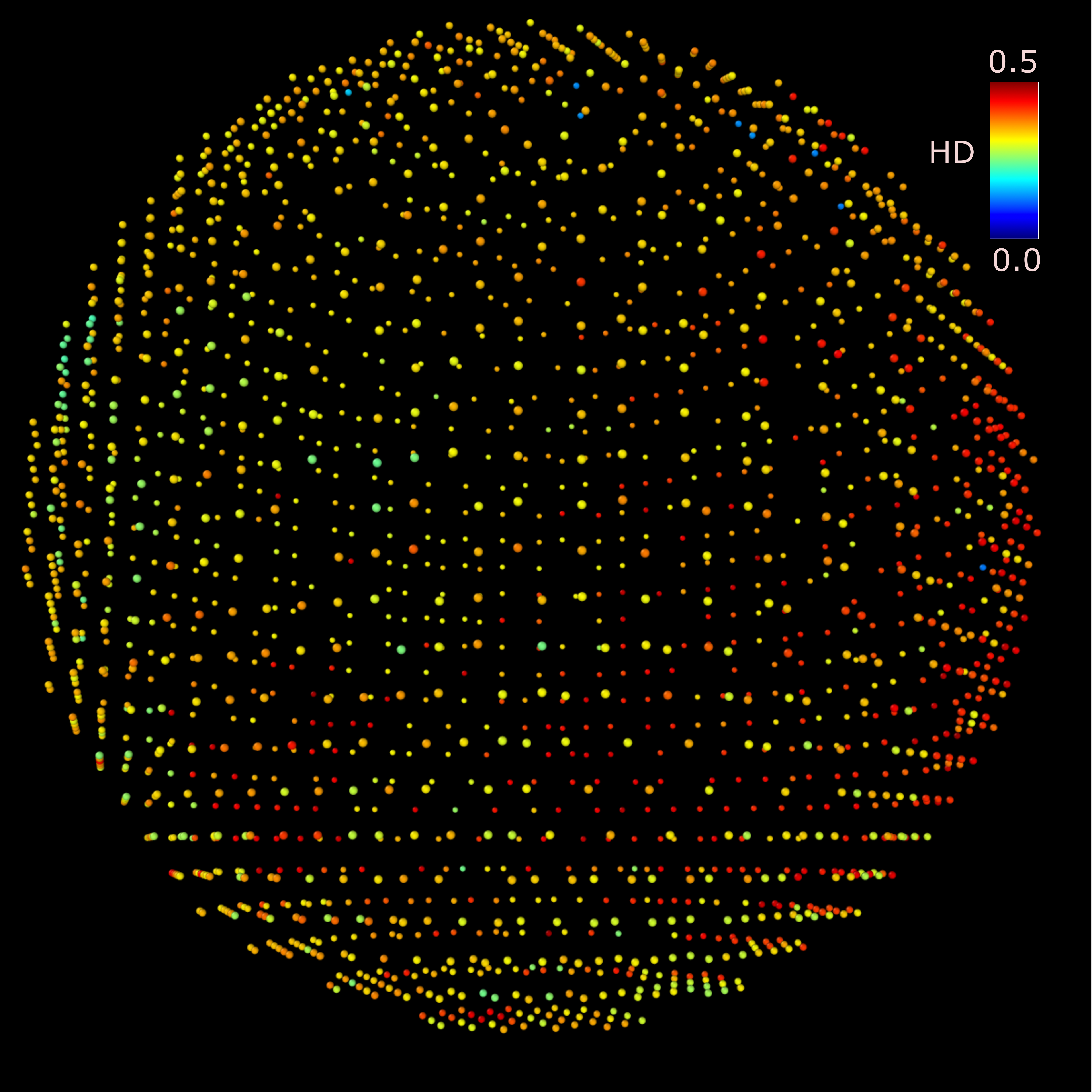}
\caption{{\bf Visual depiction of the heterogeneity. The image was generated using 3D lattice simulations with $\alpha=0.55$ and $p=0.003$. The image corresponds to tumor evolution at $t = 113$ generation}. Figure shows the HD values for $2,000$ cells, depicted by their color, located approximately $15$ lattice units from the center. The color bar, on the top right corner, shows the $HD$ scale. The large variation in the sub-population is evident.  }
\label{pic_het}
\end{figure}

\clearpage

\bibliography{ith_reference}
\bibliographystyle{achemso}

\end{document}


\singlespacing

\title{Supporting Information: Statistical Mechanical theory for spatio-temporal evolution of Intra-tumor heterogeneity in cancers: Analysis of Multiregion sequencing data}
\author{Sumit Sinha$^1$, Xin Li$^2$, D. Thirumalai$^2$}
\email{dave.thirumalai@gmail.com}
\affiliation{$^1$Department of Physics, University of Texas at Austin, Austin, TX 78712, USA.}
\affiliation{$^2$Department of Chemistry, University of Texas at Austin, Austin, TX 78712, USA.}

\date{\today}

\pacs{}

\maketitle
\section{Tumor mutation burden (TMB) from M-Seq data}
\label{M-seq reveals high number of mutations in exogenous cancers}
\begin{figure}[h]
\includegraphics[width=15cm]{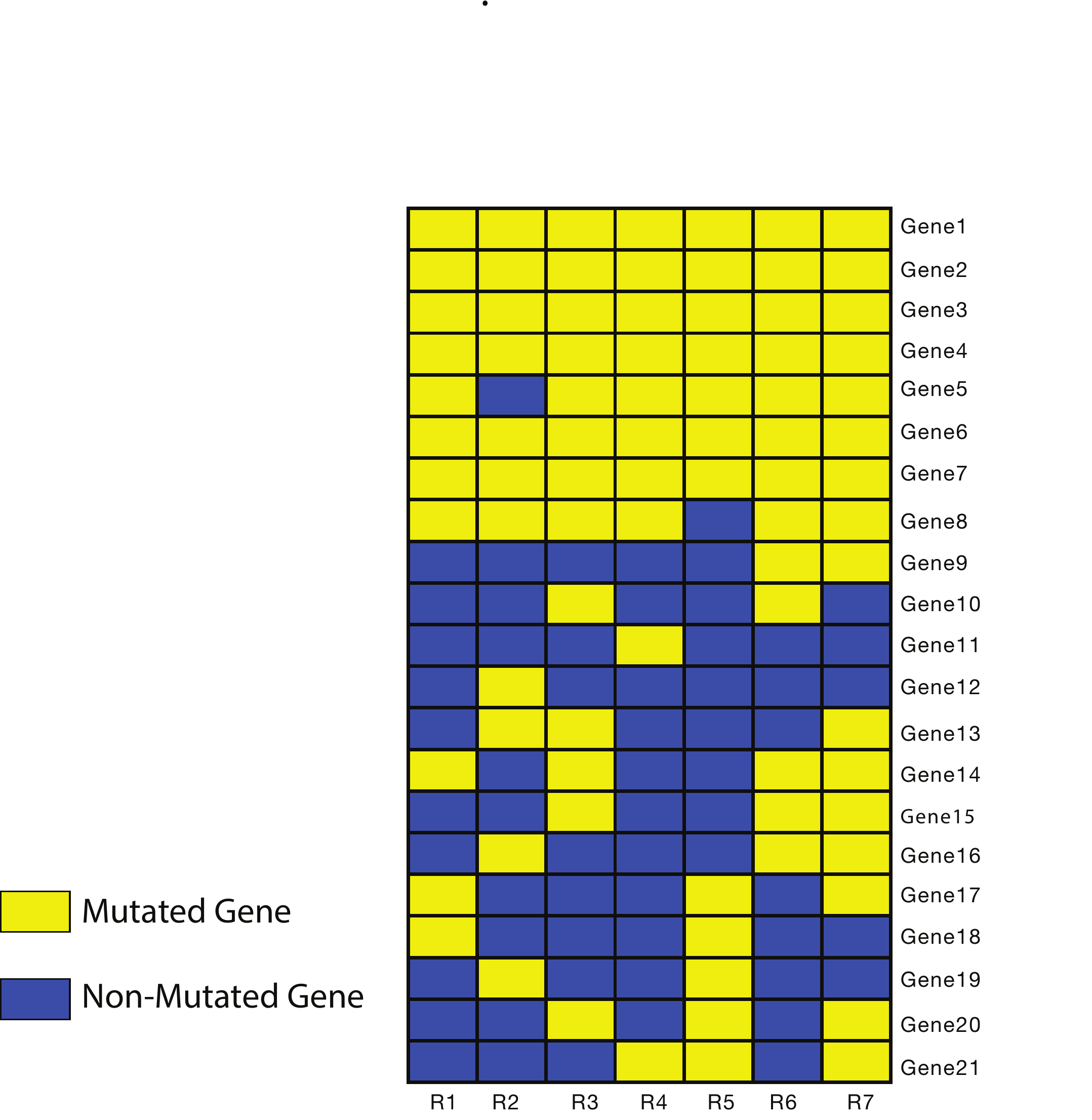}
\caption{Schematic matrix representation of a typical mutiregion sequencing (M-seq) data depicting region-specific distribution of mutations within a single tumor. The rows denote the gene names ( Gene1 to Gene21), and the columns are the regions of the sequenced tumor ( R1 to R7). The yellow (blue) boxes represent a mutated (non-mutated) genes. In the experiments, the number of regions sequenced may vary depending on the size and quality of the tumor sample. The number of genes mutated also changes across various cancer types. Note that mutations in Gene 1, Gene 2, Gene 3 are public, the mutations in Gene 11 is private, and an example of branched mutation is Gene 20.   }
\label{MRS}
\end{figure}

Multi-region sequencing (M-seq), which has provided the most direct evidence for pervasive signature of Intratumor heterogeneity (ITH) in solid tumors, is used to delineate the distribution of mutations in different regions of a single solid tumor \cite{gerlinger2012intratumor, gerlinger2014, harbst2016, debruin2014, zhang2014intratumor, cao2015}. Figure \ref{MRS} shows a schematic representation of a typical M-seq data. The data is in the form of a matrix where the rows are the names of the genes and the columns label the biopsied tumor regions. In Figure \ref{MRS}, the mutated genes are in yellow, and blue denotes the non-mutated genes. Usually the mutations, sequenced using M-seq, are non-silent or non-synonymous \cite{harbst2016,debruin2014, cao2015,gerlinger2014}, which implies that they impact the tumor phenotype. From the M-seq data, we can classify mutations as public, branched or private. Public genes are mutated in all the sequenced regions. If a mutation is present in more than one region but not in all, then it is referred to as branched. If a mutated gene is found in only one region, then it is classified as private \cite{sottoriva2015big}. The presence of a large number of branched and private mutations, enhances ITH. 

 We first analyzed the M-seq data for four cancer types from different patients in order to assess the extent of ITH. Three of them (skin, lung and esophagus) are predominantly caused by exogenous factors while the fourth ( kidney) is linked to endogenous factors \cite{tomasetti2017stem}. The data for skin cancer was taken from Harbst et al.\cite{harbst2016}, lung from Debruin et al.\cite{debruin2014}, esophagus from Cao et al.\cite{cao2015}. For kidney cancer, we used data from the experiments by Gerlinger et al.\cite{gerlinger2014}. The number of patients, whose tumors were sequenced, are eight\cite{harbst2016}, seven\cite{debruin2014}, two\cite{cao2015} and ten\cite{gerlinger2014} for skin, lung, esophagus and kidney cancer respectively. In some sense, the sample sizes are small but are sufficient to calculate ITH using our theory, and establish its efficacy.

\begin{figure}[h]
\includegraphics[width=15cm]{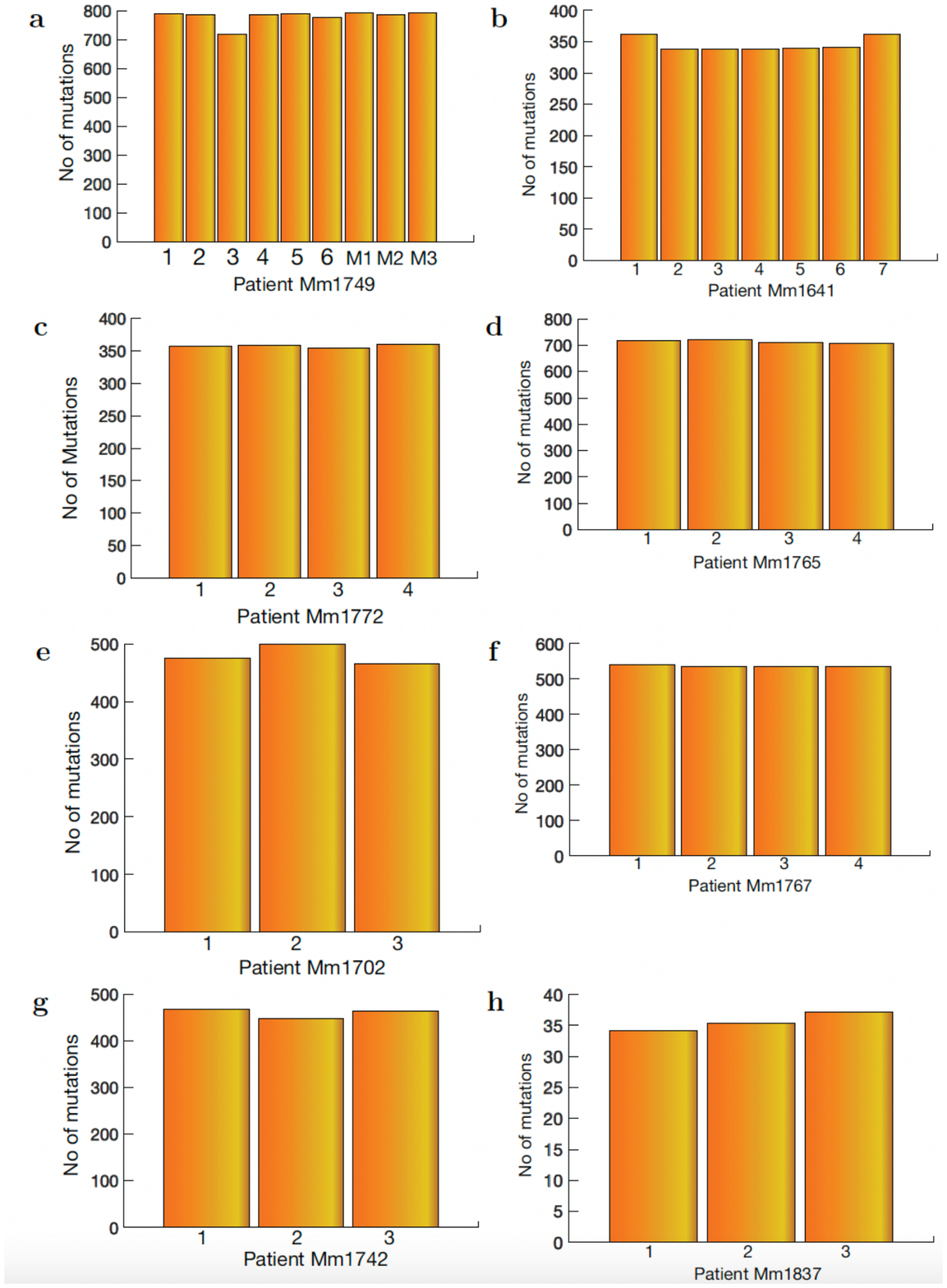}
\caption{Number of mutations in different regions of tumors for eight skin cancer patients. The number of mutations varies from patient to patient. }
\label{skincancer}
\end{figure}

\begin{figure}[h]
\includegraphics[width=15cm]{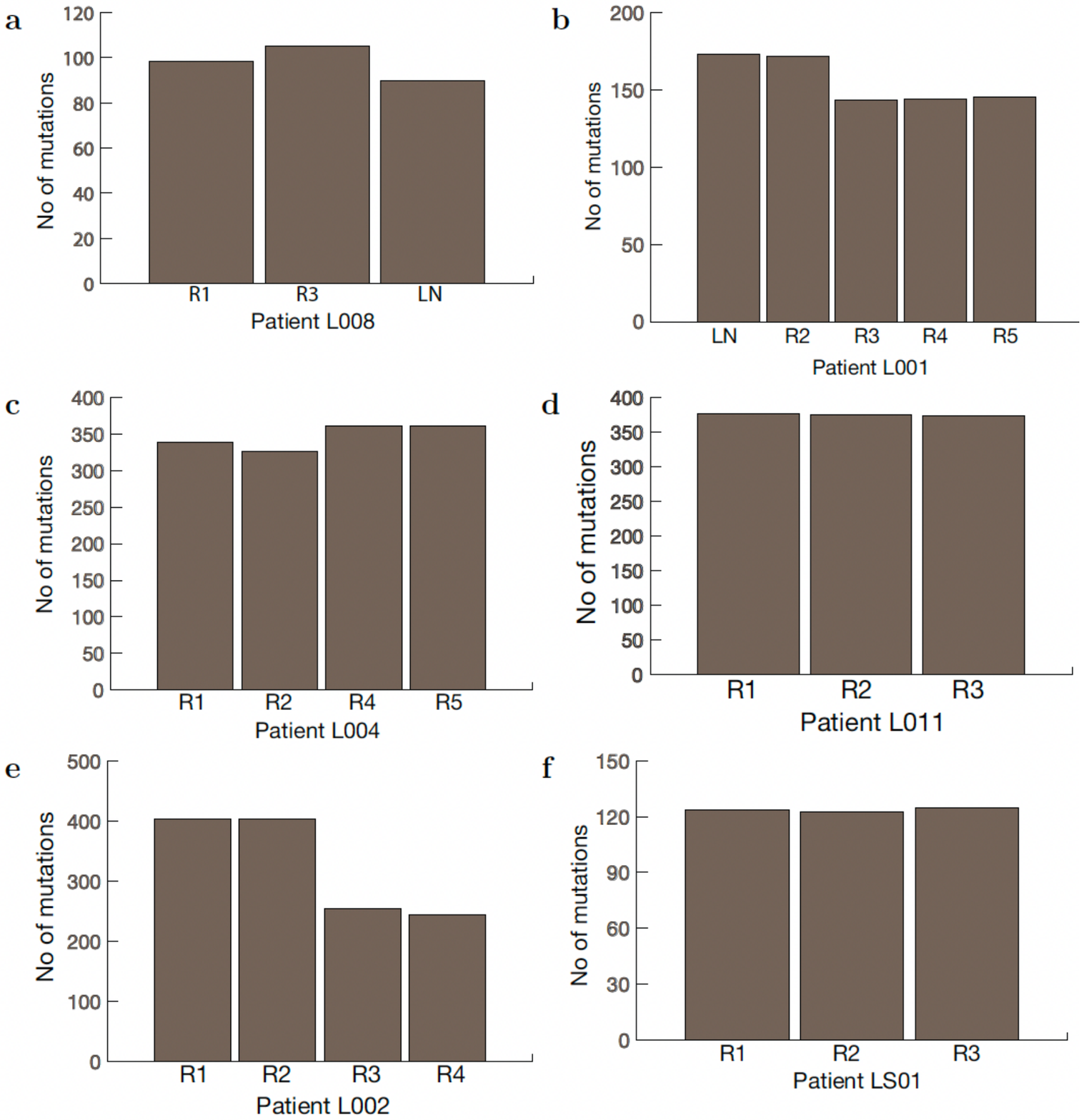}
\caption{Number of mutations in different regions of tumors for six patients with lung cancer. The average number of mutations among the six lung cancer patients was found to be $\approx 238$. The maximum number of mutations was found in patient L011 with $\approx 375$ mutations per tumor region. The minimum number of mutations was found in L008 with $\approx 98$ mutations per tumor region.}
\label{lungcancer}
\end{figure}

\begin{figure}[h]
\includegraphics[width=15cm]{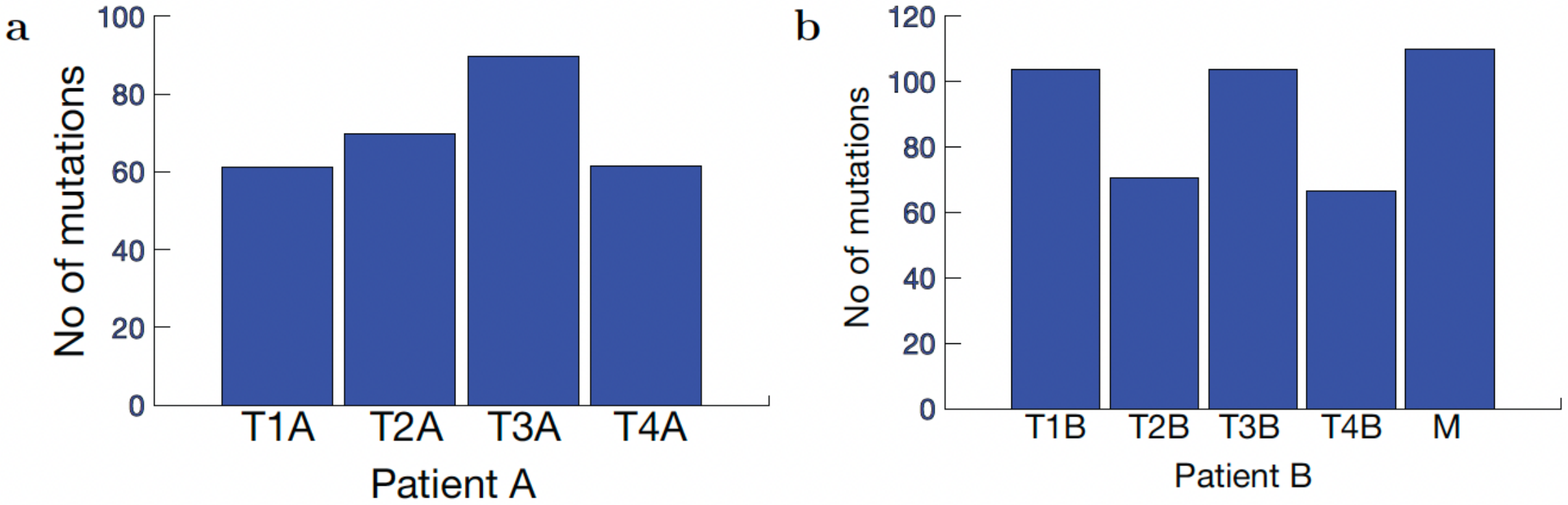}
\caption{Number of mutations in different regions of tumors for two esophageal cancer patients (A and B). Patient A had $\approx 71$ mutations per tumor region whereas patient B had $\approx 91$ mutations per tumor region.}
\label{esophagealcancer}
\end{figure}

\begin{figure}[h]
\includegraphics[width=15cm]{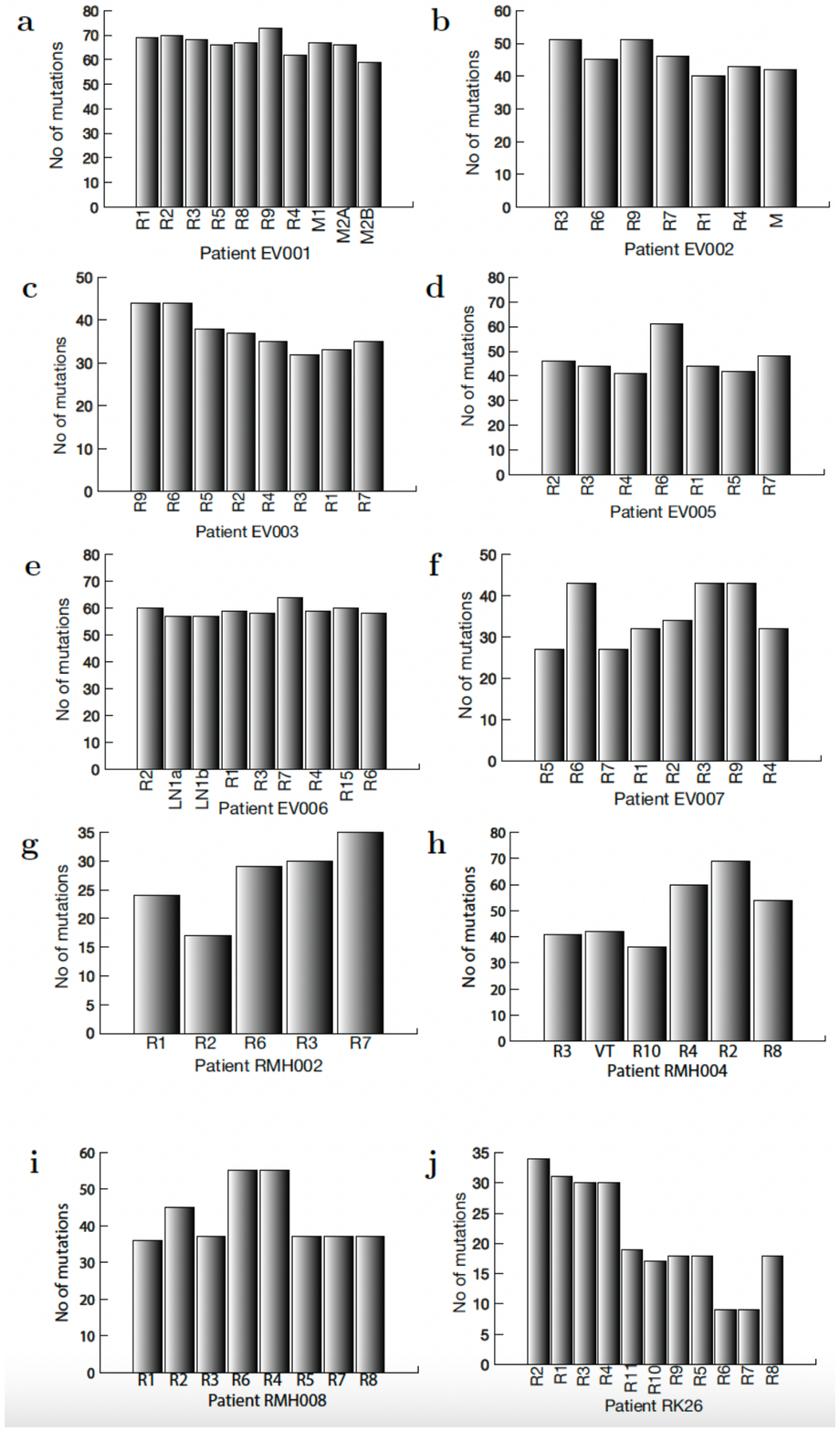}
\caption{Number of mutations in different regions of tumors across ten patients with kidney cancer. The average number of mutations across these ten patients was $\approx 43$. Patient EV001 with $\approx 67$ mutations per tumor region had the maximum whereas patient RK26 with $\approx 21$ mutations per region had the minimum number of mutations.}
\label{kidneycancer}
\end{figure}

  Figure \ref{skincancer} shows the number of mutations, referred to as tumor muation burden (TMB), in different regions of tumors across eight skin cancer patients. The average number of mutations among the eight patients is $\approx 464$. The maximum number of mutations was found in patient Mm1749 who acquired $\approx 779$ mutations per tumor region. In contrast, patient Mm1837 has the smallest number (36 per tumor region) of mutations. Figure \ref{lungcancer} summarizes the results for the six lung cancer patients. The average number of mutations among the six lung cancer patients is $\approx 238$. The maximum (minimum) number of mutations was found in patient L011 (patient L008) with $\approx 375$ ($\approx 98$) mutations per tumor region. Figure \ref{esophagealcancer} shows the number of mutations in two patients (A and B) with esophageal cancer. Patient A had $\approx 71$ mutations per tumor region whereas patient B had $\approx 91$ mutations per tumor region. Figure \ref{kidneycancer} shows the number of mutations in ten kidney cancer patients. The average number of mutations across these ten patients was $\approx 43$. Patient EV001 with $\approx 67$ mutations per tumor region had the maximum whereas patient RK26 with $\approx 21$ mutations per region had the minimum number of mutations. 

The data shows that cancers caused predominantly due to exogenous mutagens have more mutations. These results are consistent with previous studies \cite{kandoth2013mutational, lawrence2013mutational}.  However, strong claims about esophageal cancer should be made with caution because M-seq data is available only for two patients.

\section{Relationship between ITH and TMB}
The tumor mutation burden, TMB, is defined as the number of mutations accumulated over a time period, $t$. In the M-Seq data, the average number of mutations per region is $(1-x_t)n$, where $n$ is the length of the string. Because $x_t=\frac{1\pm \sqrt{1-4H'}}{2}$, one can derive a relationship between TMB and ITH using $H'=\frac{ITH}{2F(\alpha)}$. The relationship is given by $TMB=n[\frac{1\mp \sqrt{1-4H'}}{2}]$.
One can also estimate the number of single nucleotides mutated per year from the M-Seq data ($M_{SN}$) by using the relationship, 
\begin{equation}
G(1-(1-p_{D})^t)=n(1-x_t), 
\end{equation}
where $p_D$ is the mutation probability of a gene per year, G is total number of genes in the genome ($\approx 20,000$), and $t$ is the time from tumor initiation to detection. For skin and kidney cancers, $t$ is about 13, and 20 years, respectively\cite{li2021imprints}. For lung cancer, the TMB is found to show a weak correlation with the patient age, therefore, the mutation rate is not detected in Ref.~\cite{li2021imprints}. Here, we estimated the $M_{SN}$ for skin and kindey cancer patients. Using, $M_{SN}=(\frac{p_D}{N_G})(N_D)=\frac{(p_D)}{(10^{4})}(3\times 10^9)$, where $N_G$ ($\approx 10^4~bps$) is the average number of nucleotides in a gene (10,000) and $N_D$ is the number of nucleotides on the DNA. Figure \ref{m_sn} shows the plot for $M_{SN}$ for the two cancer types: skin, and kidney. For skin and kidney cancers, the mean $M_{SN}$ is $\sim 1200$ and  $\sim 125$ respectively, which is consistent with our previous findings, Skin ($\sim 1500$) and Kidney ($\sim 70$), from The Cancer Genome Atlas (TCGA) data (see Figures~1 (Kidney) and 3 (Skin) in \cite{li2021imprints}).

\section{Tumor Simulation Model}
We created a three-dimensional (3D) cellular automaton (CA) model, similar to a computational model studied previously in a pioneering paper \cite{gonzalez2002metapopulation}, to calculate ITH in order to not only validate the theory but provide mechanistic insights into the spatio-temporal propagation of ITH. The 3D space is divided into finite lattice sites of size $N \times N \times N$, with $N=100$. A lattice site, whose position is represented as $[i, j,k]$, is either vacant ($L[i, j,k]=0$) or occupied ($L[i, j,k]=1$). A lattice site can only be occupied by a one cell (see Fig~\ref{sim_des}a).

In the process of evolution, a normal cell may become cancerous by accumulating non-synonymous (or non-silent) mutations. The number of non-synonymous mutations, $n_d[i,j,k]$, at position $[i, j,k]$, characterizes whether a cell is normal or cancerous. If $n_d[i,j,k]=0$, then the cell is normal otherwise it is cancerous (see Fig.~\ref{sim_des}b). 

 We model the DNA  representing the mutation sites to calculate the ITH. Each cell contains DNA in the form of an information string, $X$, of length $n$. A site in $X$ is $0$ (no mutation) or $1$ (mutated). For simplicity, we consider the same fitness advantage, $s_d$, for all the non-synonymous mutations. We assume $s_d=0$, unless stated otherwise. The assumption that $s_d=0$ implies that evolution is neutral. As a result, there is no fitness advantage upon mutation.

Cell division is controlled by birth probability ($\alpha$) provided neighboring lattice site (see Fig.~\ref{sim_des}c) is vacant. In 3D, the neighbors are defined by the 26 lattice sites around a cell. The birth probability of a cell, $\alpha[i,j,k]$, is given by the relation,
 \begin{equation}
 \alpha[i,j,k]=\alpha_o[i,j,k] + s_d \times n_d[i, j,k],
 \label{eq:fitness}
  \end{equation}
where, $\alpha_o[i,j,k] $ is the initial birth probability of the cell. When a cell divides, we copy the mutation profile of the parent cell ($P$) to its daughter cell (D) ( $X^{P}=X^{D}$).

 At each time step (equivalent to one cell generation), a cell may acquire non-synonymous mutations with probability $p$ per site. We also assume that there is no back mutation. A cell can also undergo apoptosis with  probability $\beta[i, j,k]= 1-\alpha[i,j,k]$ (see Fig.~\ref{sim_des}d).
 
 {\bf Evolution with selection:} To assess the robustness of the conclusion that ITH is pervasive and varies greatly within a single tumor, we also carried out simulations with $s_d \ne 0$. The effect of fitness advantage is modeled by increasing the birth rate given in \ref{eq:fitness}. Figure \ref{fig:fitness_ad}, shows that $s_d$ plays a role in the range, $0.3\leq x_t \leq 0.7$. For $s_d =0.0001$ and $0.001$, we see that the ITH remains approximately the same as in the case of $s_d=0$. However, for the case of $s_d=0.01$, we see that heterogeneity increases slightly in the domain $0.3\leq x_t \leq 0.7$. This occurs because as we increase $s_d$, the birth probability increases as a function of time, and hence it will result enhance heterogeneity. We know that at higher birth probability, there is more heterogeneity as explained in the main text.

{\bf Initial Conditions:} To initiate the simulations, we introduce a single normal cell at the center of the 3D lattice. This cell has a birth probability  $\alpha[\frac{N}{2},\frac{N}{2},\frac{N}{2}]=\alpha_o$ without any non-synonymous mutations ($n_d[\frac{N}{2},\frac{N}{2},\frac{N}{2}] =0$). Following the evolutionary rules described above, this cell divides and acquires mutations stochastically resulting in a tumor mass with heterogeneous DNA strings. 

{\bf Termination of Simulations: }We stop the simulations once the size of the tumor reaches $M$ cells. We assume that $M=50,000$, if not mentioned explicitly. The simulations were performed using an in house code using the MATLAB software.

\begin{table}[ht]
\centering 
\caption{Hamming distance (HD) between the pairs of columns for the data in Figure 1b of the main text. }
\scalebox{0.8}{\begin{tabular}{|c|c| } 
\hline
Pairs & Hamming Distance(HD) \\ [0.5ex] 
\hline 
(R1, R2) &  0.13 \\ 
\hline 
(R1, R3) & 0.10  \\
\hline 
(R1, R4) & 0.07  \\
\hline 
(R1, R5) & 0.04  \\
\hline 
(R1, R6) & 0.31 \\ 
\hline 
(R1, R7) & 0.18 \\
\hline 
(R2, R3) & 0.08 \\
\hline 
(R2, R4) & 0.06 \\
\hline 
(R2, R5) & 0.12 \\
\hline 
(R2, R6) & 0.30 \\
\hline 
(R2, R7) & 0.22 \\
\hline 
(R3,R4) & 0.02\\
\hline 
(R3, R5) & 0.08\\
\hline 
(R3, R6) & 0.27\\
\hline 
(R3, R7) & 0.18\\
\hline 
(R4, R5) & 0.06\\
\hline 
(R4, R6) & 0.24\\
\hline 
(R4, R7) & 0.16\\
\hline 
(R5, R6) & 0.30\\
\hline 
(R5, R7) & 0.17\\
\hline 
(R6, R7) & 0.40\\ [1ex] 
\hline 
\end{tabular}}
\label{tableone} 
\end{table}

\clearpage

\clearpage

\clearpage

\clearpage

\clearpage

\clearpage
\begin{figure}[h]
\includegraphics[width=15cm]{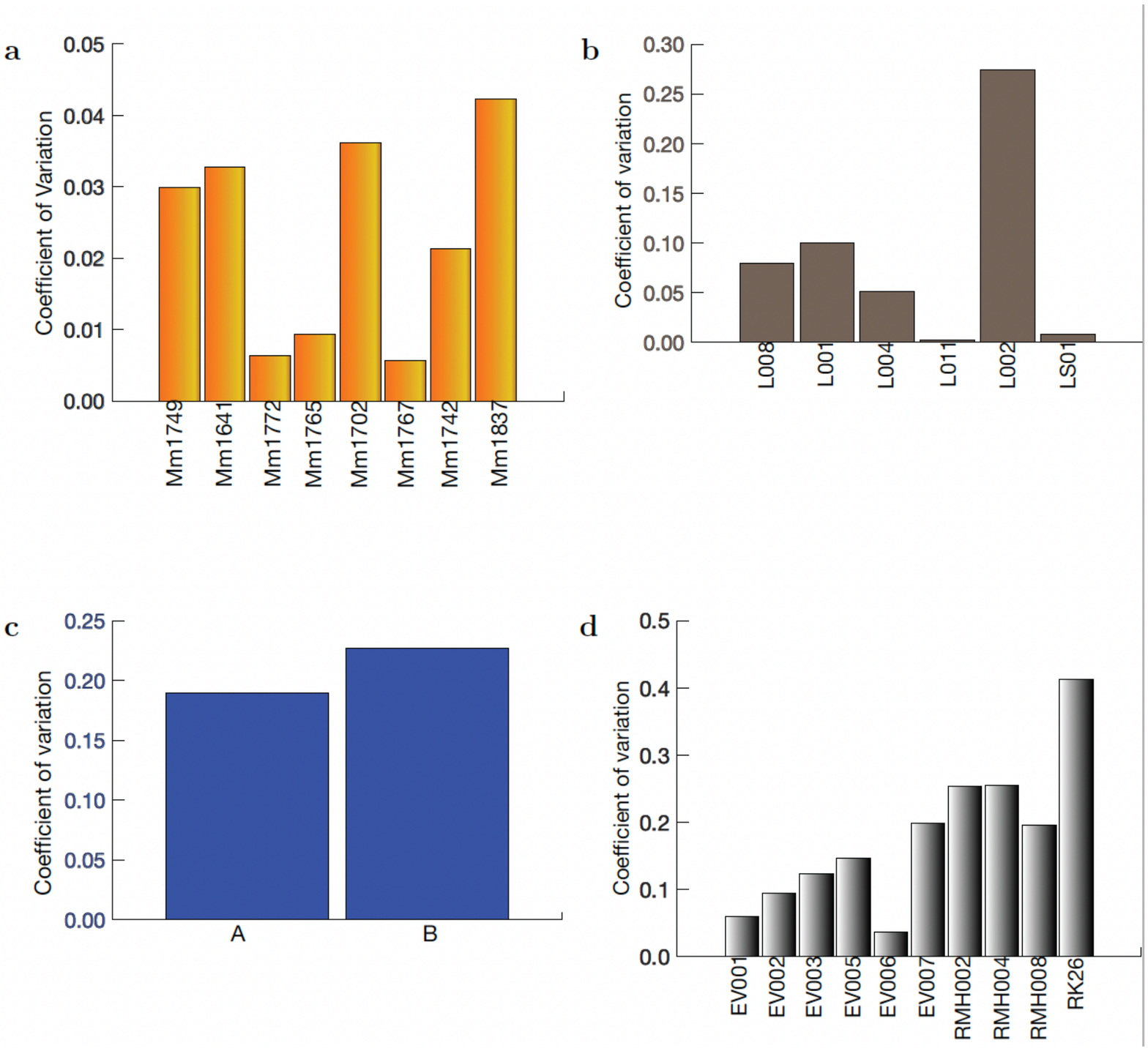}
\caption{Coefficient of variation ($c_v$), defined in Eq.(10) in the main text, for exogenous (skin, lung and esophagus) and endogenous (kidney) cancers. \textbf{(a)} $c_v$ for eight skin cancer patients. The $c_v$ varies from 0.042 (patient Mm1837) to 0.006 (patient Mm1767). \textbf{(b)} Same as (a) except for six lung cancer patients. $c_v$ varies from $\approx 0.002$ (patient L011) to $\approx 0.27$ (patient L002). \textbf{(c)} shows the $c_v$ for the two esophageal cancer patients (A and B). \textbf{(d)} Same as (a) except for the ten kidney cancer patients. In this case, $c_v$ changes from $\approx 0.004$ (patient EV006) to 0.4 (patient RK26).}
\label{cv}
\end{figure}

\clearpage
\begin{figure}[h]
\includegraphics[width=12cm]{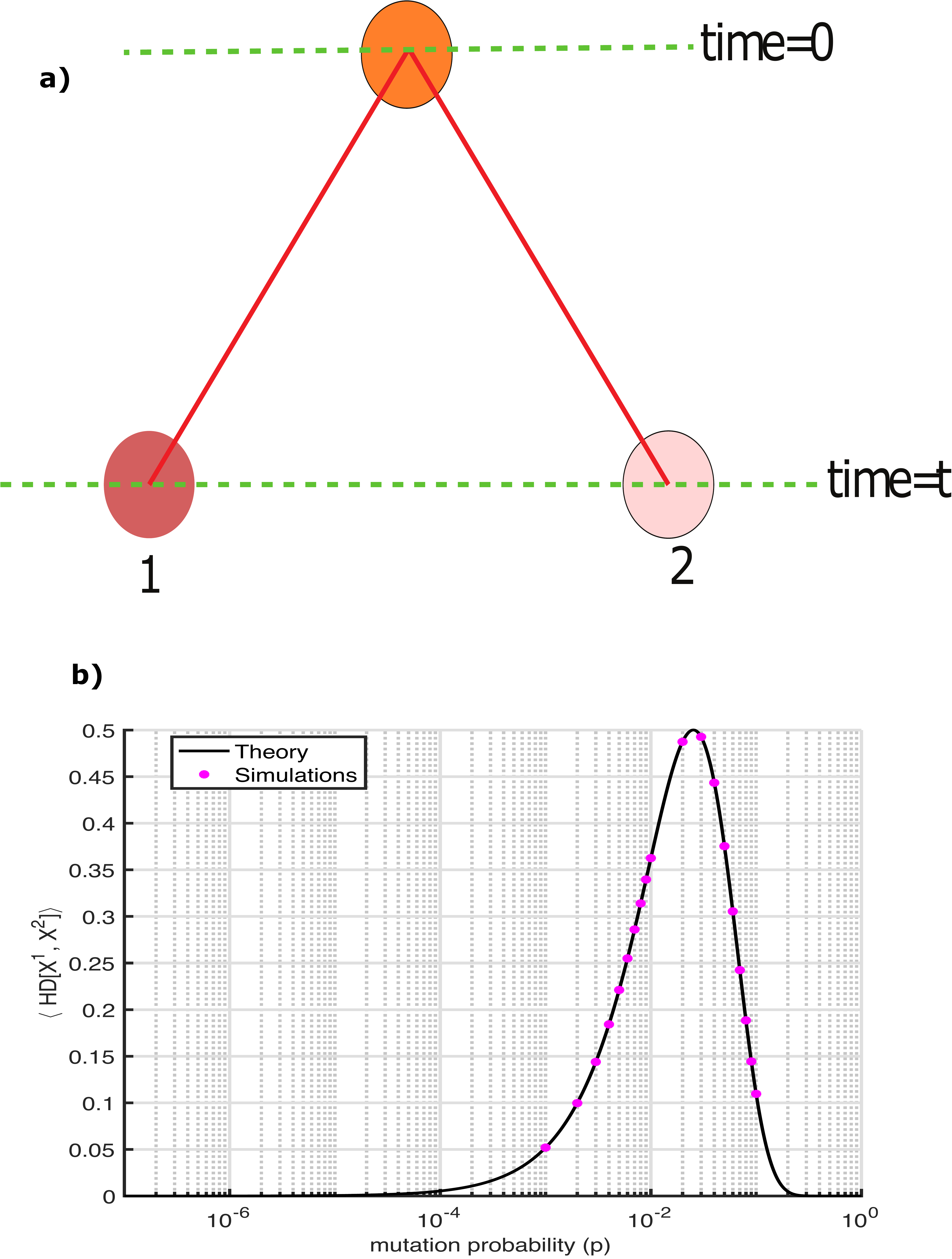}
\caption{{\bf Independent Evolution:} {\bf (a)} Evolution of two cells (brown and pink). The two cells are normal at time $t=0$ (orange), and evolve independently acquiring exogenous mutations. {\bf (b)} The average heterogeneity between the two cells ($\langle HD[X^1,X^2]\rangle$) that evolve independently as a function of the mutation probability (p). The dots are simulation results carried out for $27$ time steps or generations. The simulation was averaged over $\approx 4\times 10^8$ pairs of cells. The black line is calculated using the theoretical prediction given in Eq. (18) in the main text.}
\label{independent}
\end{figure}

\clearpage
\begin{figure}[h]
\includegraphics[width=18cm]{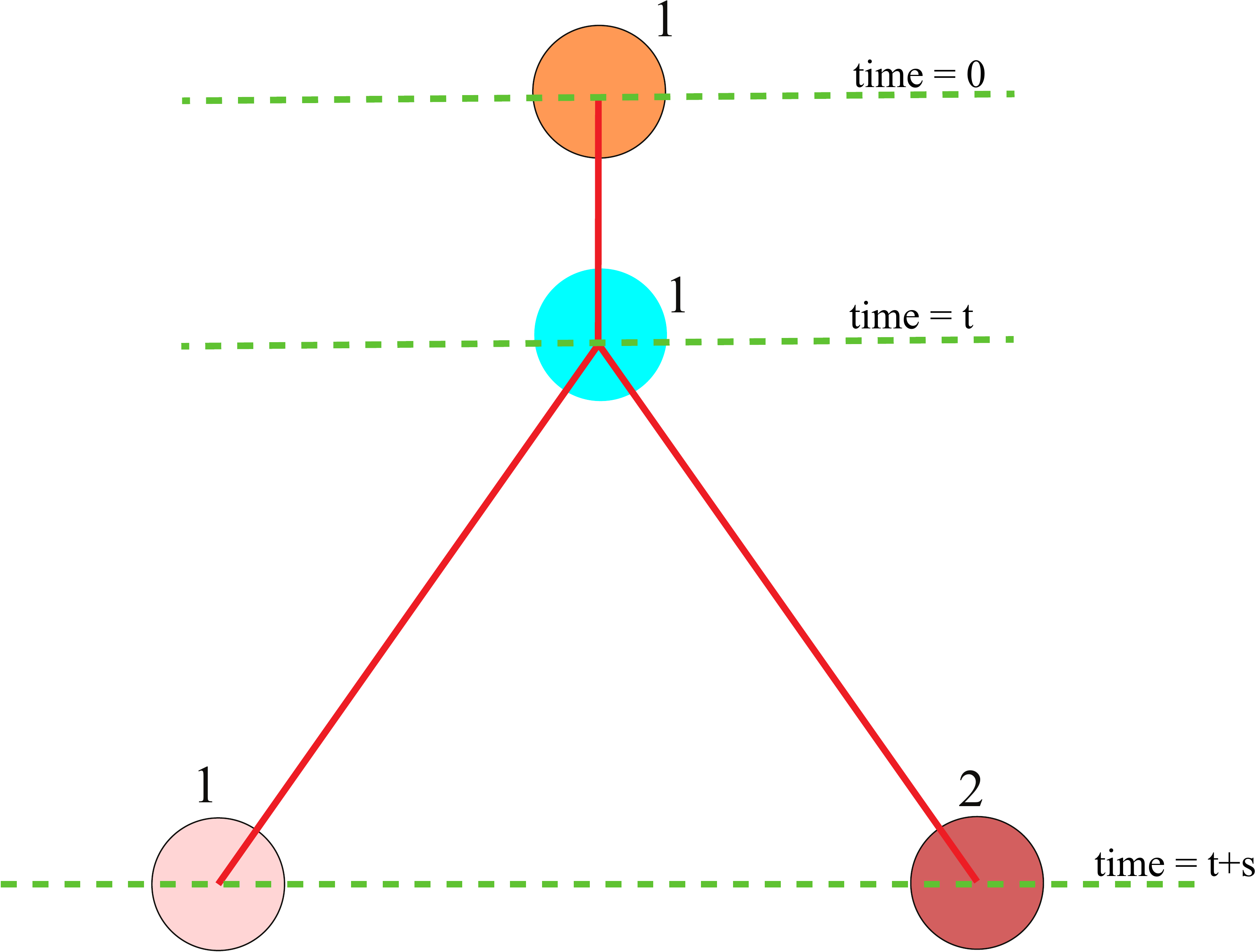}
\caption{{\bf Schematic of correlated (dependent) evolution of 2 cells (black filled circles)}. At time $t=0$, evolution is  initiated with a single normal cell (cell $1$). It evolved till time $t$, gathering exogenous mutations. At time $t$, cell $1$ gives birth to cell $2$. After time $t$, the 2 cells evolve independently for the next $s$ time steps. The ancestor of cell 2, at $t+s$, is cell 1. Therefore, we consider them to have undergone correlated evolution. At $t$, the genetic information in the blue cell is copied to the pink and brown cells. Because the pink cell at $t = t+s$ is genotypically identical to the blue cell, we use the same label.   }
\label{dependent}
\end{figure}

\clearpage
\begin{figure}[h]
\includegraphics[width=18cm]{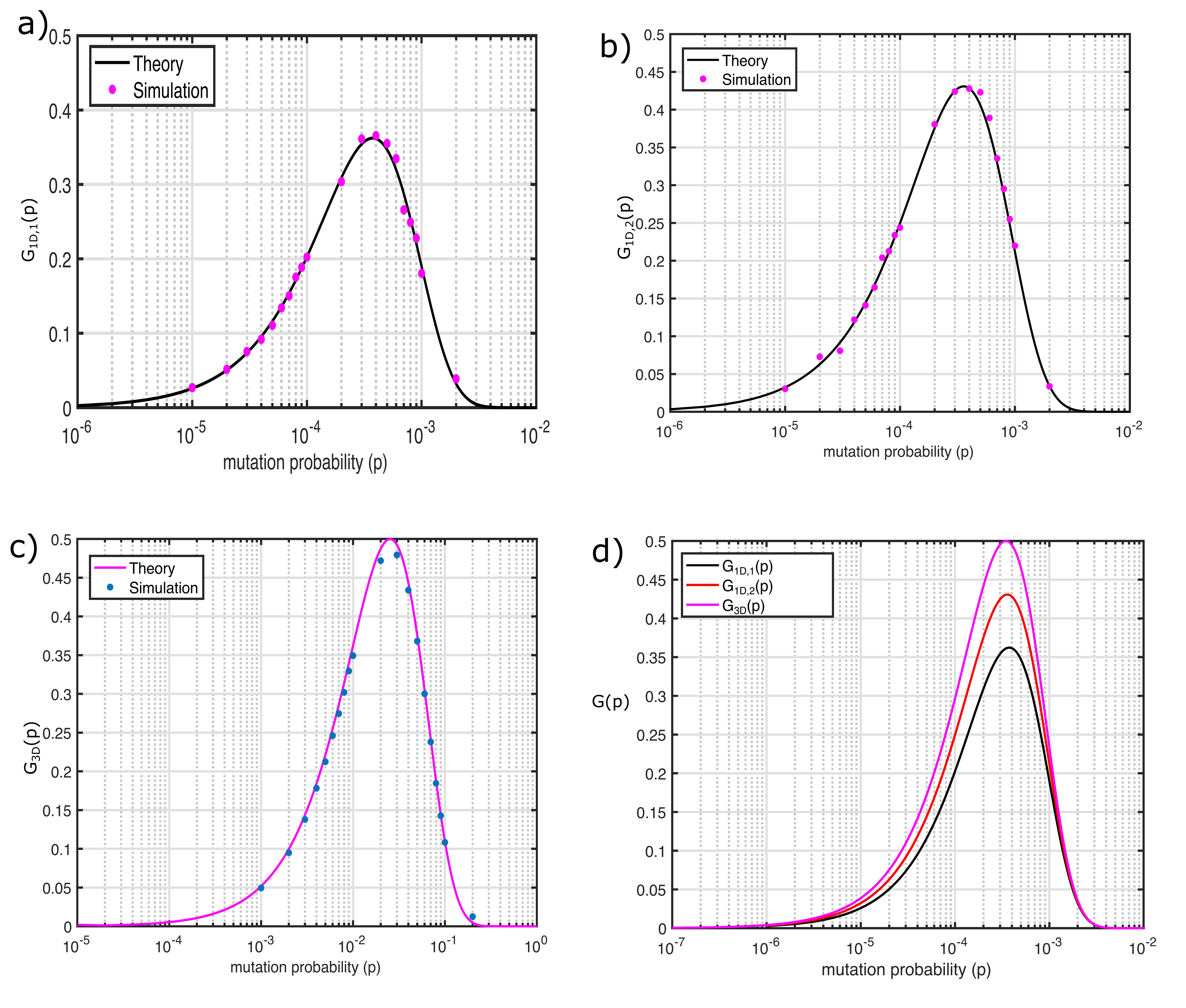}
\caption{(Continued on the next page)}
\label{g_cal}
\end{figure}

\clearpage
\begin{figure}
\contcaption{{\bf Comparison between theory and simulations for G(p) as a function of $p$ for different geometries.} In all cases $\alpha=1$, dots are results from simulations, and lines are theoretical predictions. Eqns. (25), (26) and (28) in the main text, plotted on the y-axis are evaluated at $\tau_1=2,000$ steps (generations) in (a) and (b) and $\tau_2=27$ generations in (c)). For clarity we do not show the time argument in G(p, t). {\bf (a)}Average heterogeneity within an evolving tumor in the 1D semi-infinite lattice, $G_{1D,1}(p)$, for $\alpha=1$. The dots are generated using simulations with $\alpha=1$. The number of cells at the end of simulations is $2,000$. The black line corresponds to equation (25) in the main text. Due to the correlated nature of evolution, the heterogeneity is smaller compared to the independent evolution of cells ( Figure \ref{independent}b). {\bf (b)} Average heterogeneity within an evolving tumor in 1D infinite lattice, $G_{2D,1}(p)$. The number of cells at the end of the imulation is $4,000$. The black line is calculated using Eq. (26) in the main text. Due to temporal correlations present among the cells, the heterogeneity in this case is smaller compared to the case of independent evolution of cells (i.e Figure \ref{independent}b). {\bf (c)} Same as (b) except the results for a tumor in 3D lattice, $G_{3D}(p)$. The number of cells at the end of the simulation was $\approx 50,000$. Magenta line is a plot of Eq. (28) in the main text. The behavior is similar to cells evolving independently because in 3D, the number of branches in an evolutionary tree is very large ($\propto A_c$) where $A_c$ surface area of the tumor. {\bf (d)} Comparison of average heterogeneity, $G(p)$, in 1D semi-infinite (black), 1D infinite (red) and 3D lattice (magenta). $G_{1D,1}(p)$ has the smallest value, $G_{1D,2}(p)$ has an intermediate value and $G_{3D}$ for the 3D lattice has the maximum value. The plots illustrate that heterogeneity increases as the degree of branching within an evolutionary tree increases.}
\end{figure}

\clearpage
\begin{figure}[h]
\includegraphics[width=18cm]{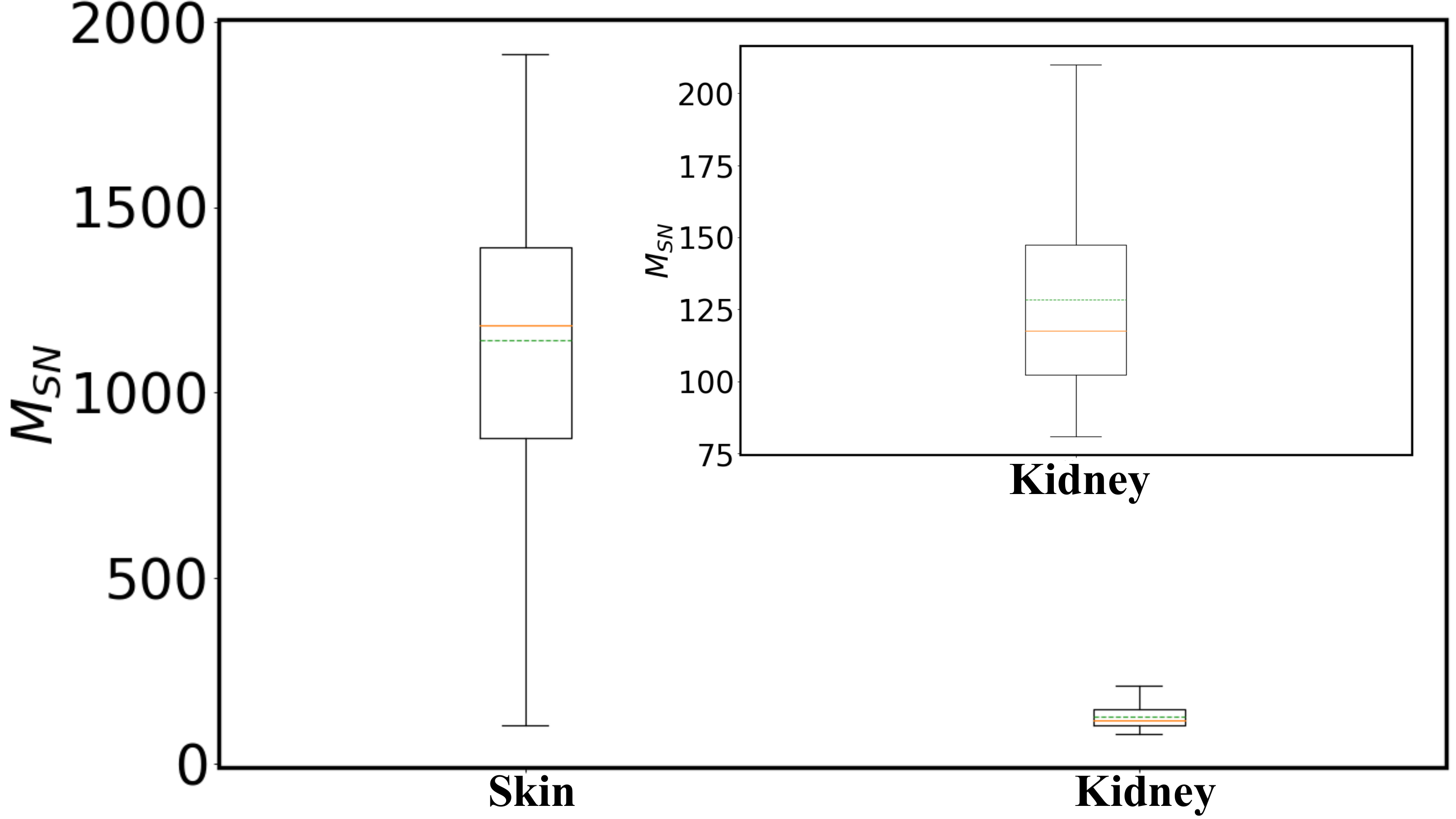}
\caption{Number of single nucleotide mutations per year ($M_{SN}$) for the skin, kidney cancer. The yellow (green) line is the median(mean) of the data. The inset shows the box plot for kidney cancer.}
\label{m_sn}
\end{figure}

\clearpage
\begin{figure}
\includegraphics[clip,width=1\textwidth]{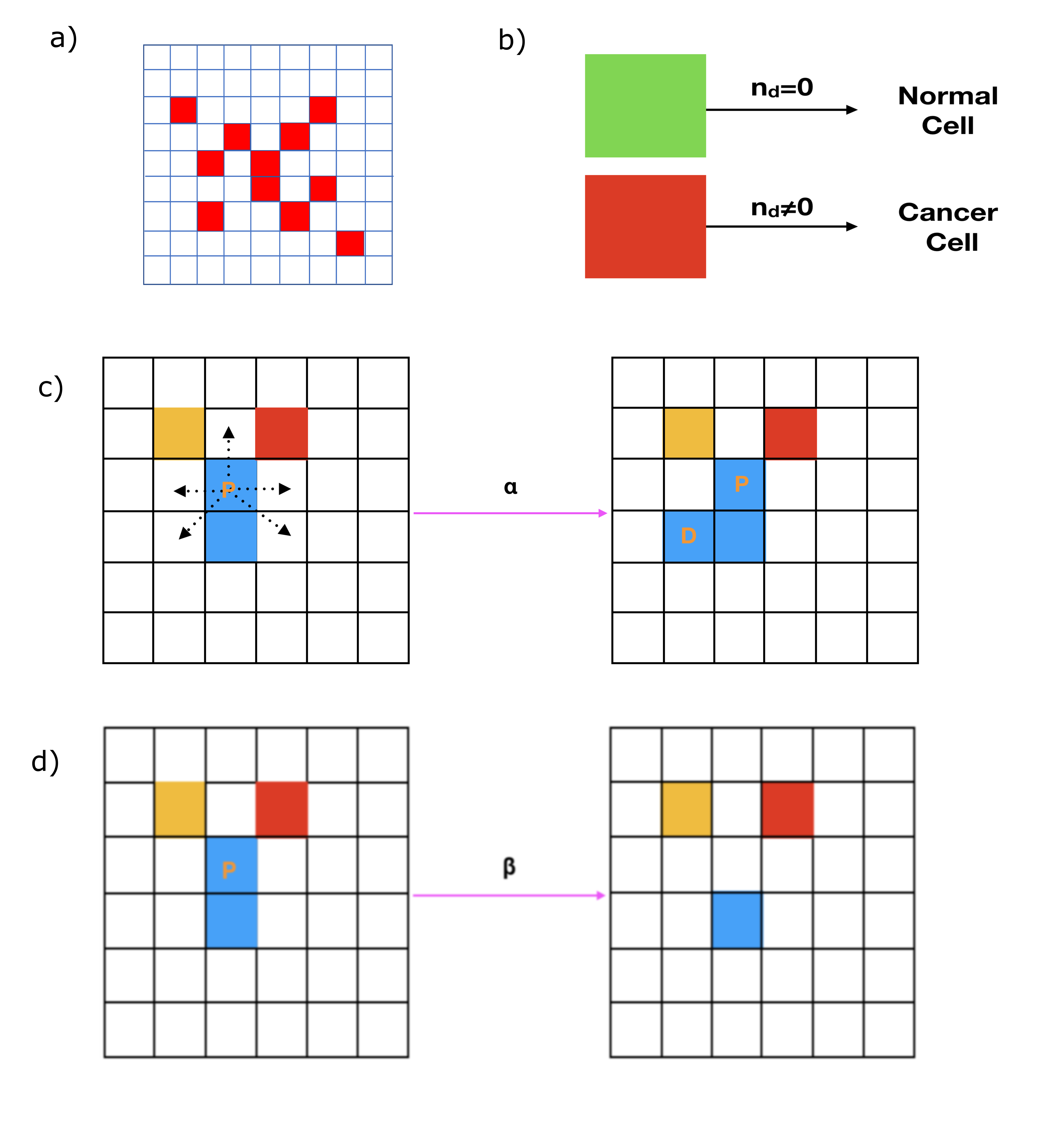}
\caption{{\bf Schematic of the simulation framework}. {\bf (a)} Cross-section of a 3D ($N^3$) lattice model for cancer evolution. For purposes of illustration, we choose $N=9$. The red colored squares denote an occupied site ($L[i, j,k]=1$) and white squares denote the site is vacant ($L[i, j,k]=0$).{\bf (b)} Cartoon depicting a cancer cell and a normal cell. A cell with zero number of non-synonymous mutations ($n_d=0$) is normal. It is cancerous otherwise ($n_d > 0$). (c),(d) Cartoon depicting cell division and apoptosis. \textbf{(c)} Shows that the parent cell (P) divides (with probability $\alpha[i, j, k]$), and a daughter cell (D) is born. The D cell occupies one of the available sites (indicated by the five dashed arrows). In \textbf{(d)} the parent cell (P) undergoes apoptosis (with a probability $\beta[i,j, k] = 1-\alpha(i,j,k)$), and the site becomes vacant. In both the figures, different colors means that cells have different genetic composition.}
\label{sim_des}
\end{figure}

\clearpage
\begin{figure}
\includegraphics[clip,width=1\textwidth]{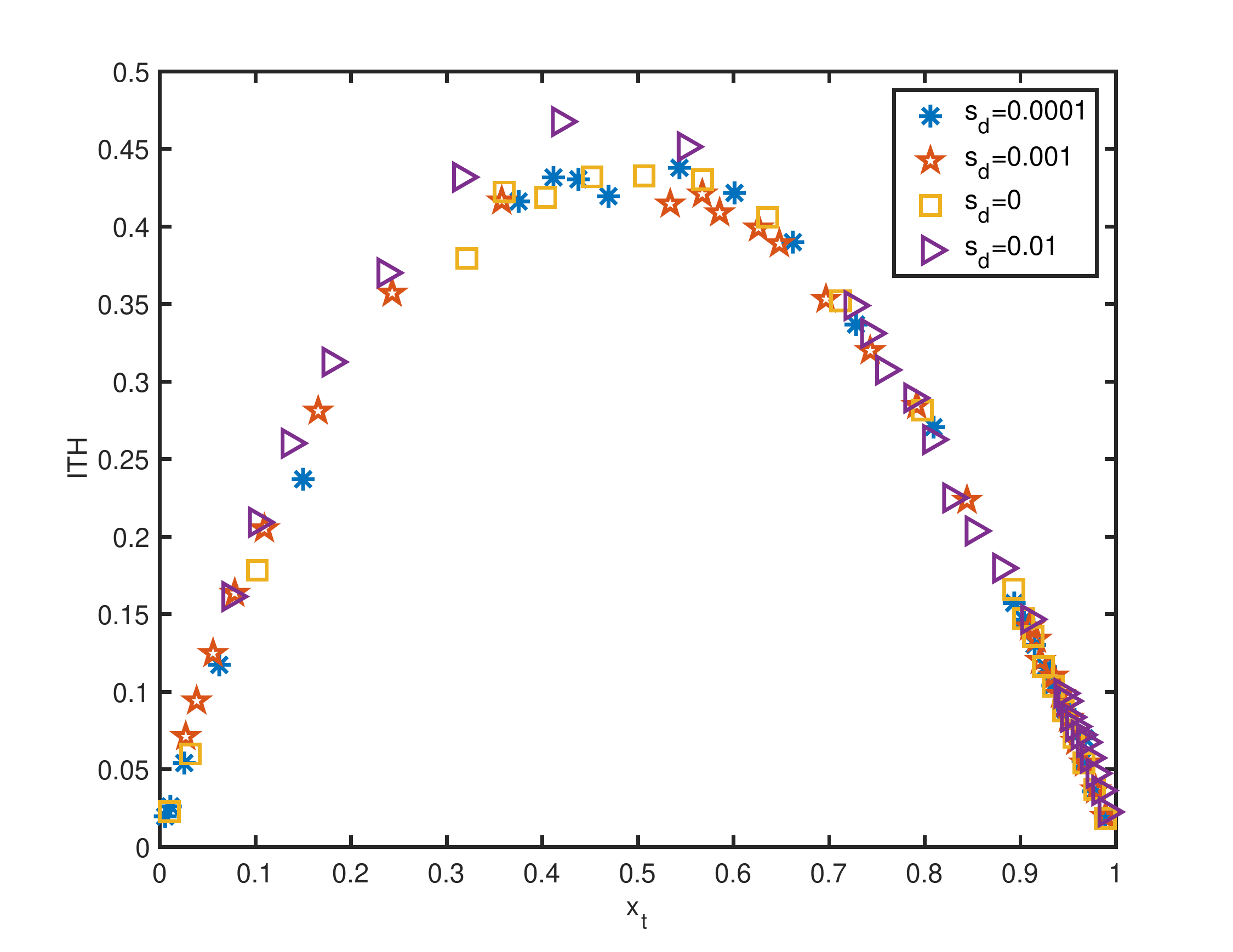}
\caption{{\bf Role of fitness advantage ($s_d$) on ITH}. ITH as a function of $x_t$ for three non-zero $s_d$ values. For comparison, we also plot the results for $s_d=0$. For $s_d=0.0001$ and $s_d=0.001$, the ITH is approximately similar to case of $s_d=0$. However, for $s_d=0.01$, we see noticeable deviations from the master curve in the range $0.3\leq x_t \leq 0.7$. The results were obtained for $\alpha=0.55$ and $t$ when the number of cells is $~50,000$.}
\label{fig:fitness_ad}
\end{figure}

\clearpage
\begin{figure}
\includegraphics[clip,width=1\textwidth]{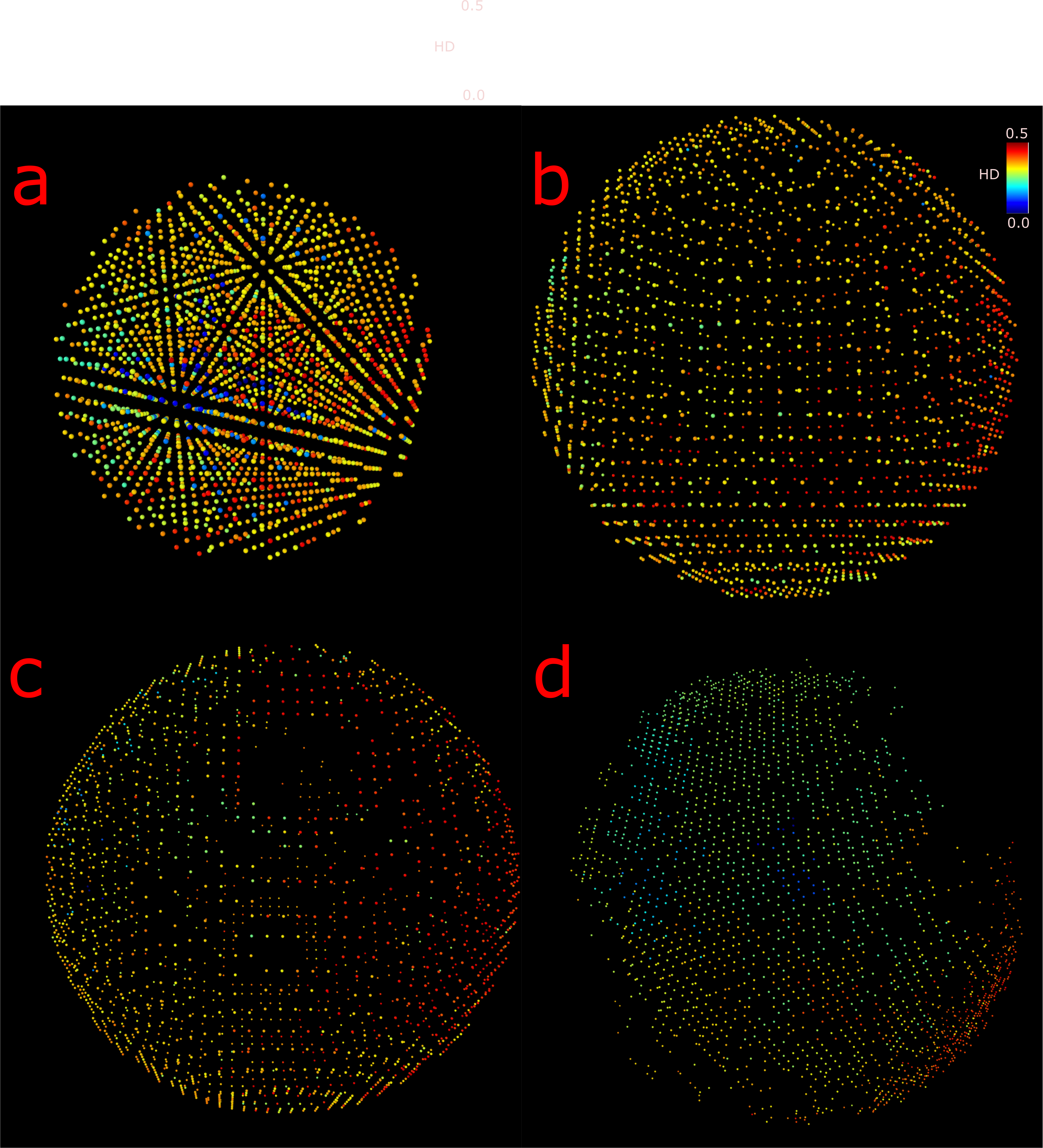}
\caption{{\bf Sub-sample to sub-sample variation in ITH in different regions of a simulated tumor}. The images were generated using 3D lattice simulations with $\alpha = 0.55$ and $p = 0.003$. The image corresponds to tumor evolution that was terminated after $t = 113$ generations. In all the cases, 2, 000 cells are present and the HD value is calculated with respect to the cell in the center. {\bf (a)} HD values for $2,000$ cells located closest to the center, depicted by their color. {\bf (b)} Same as (a) except for cells located approximately $15$ lattice units from the center (same as Figure 8 in main text). {\bf (c)} Same as (a) except for cells located approximately $20$ lattice units from the center. {\bf (d)} Same as (a) except for cells located approximately $27$ lattice units from the center. The scale on the top right corner gives the scale for HD.}
\label{fig:subsample}
\end{figure}

\clearpage

\begin{figure}
\includegraphics[clip,width=1\textwidth]{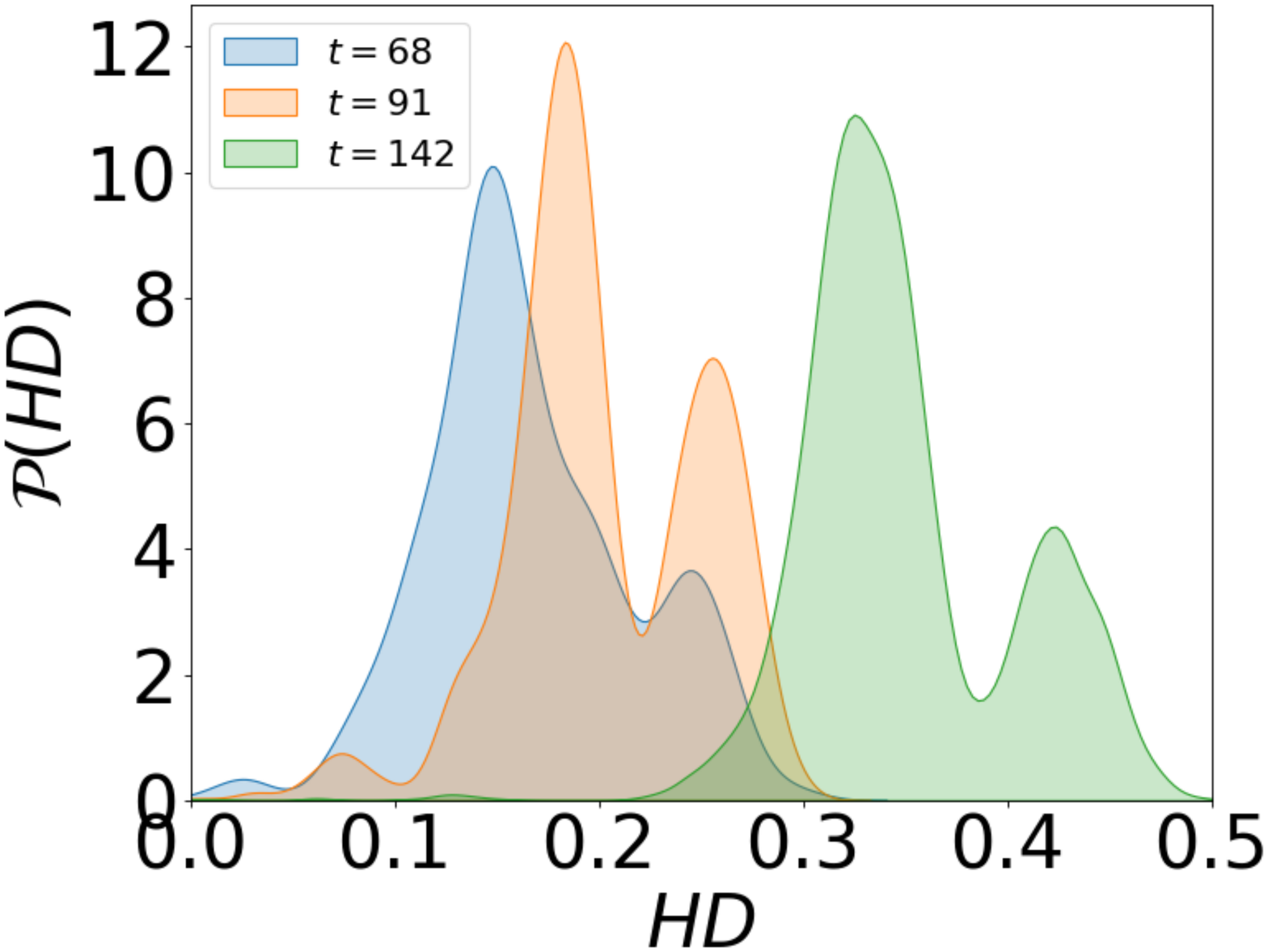}
\caption{{\bf Time evolution of the Hamming distance probability distribution $\mathcal{P}(HD)$}. From left to right, $\mathcal{P}(HD)$ corresponds to $t = 68, 91$ and $142$ generations. All the distribution were generated using 3D lattice simulations with $\alpha = 0.55$ and $p = 0.003$. HD values were calculated with respect to the center cell.}
\label{fig:subsample}
\end{figure}

\clearpage

\begin{figure}
\includegraphics[clip,width=0.85\textwidth]{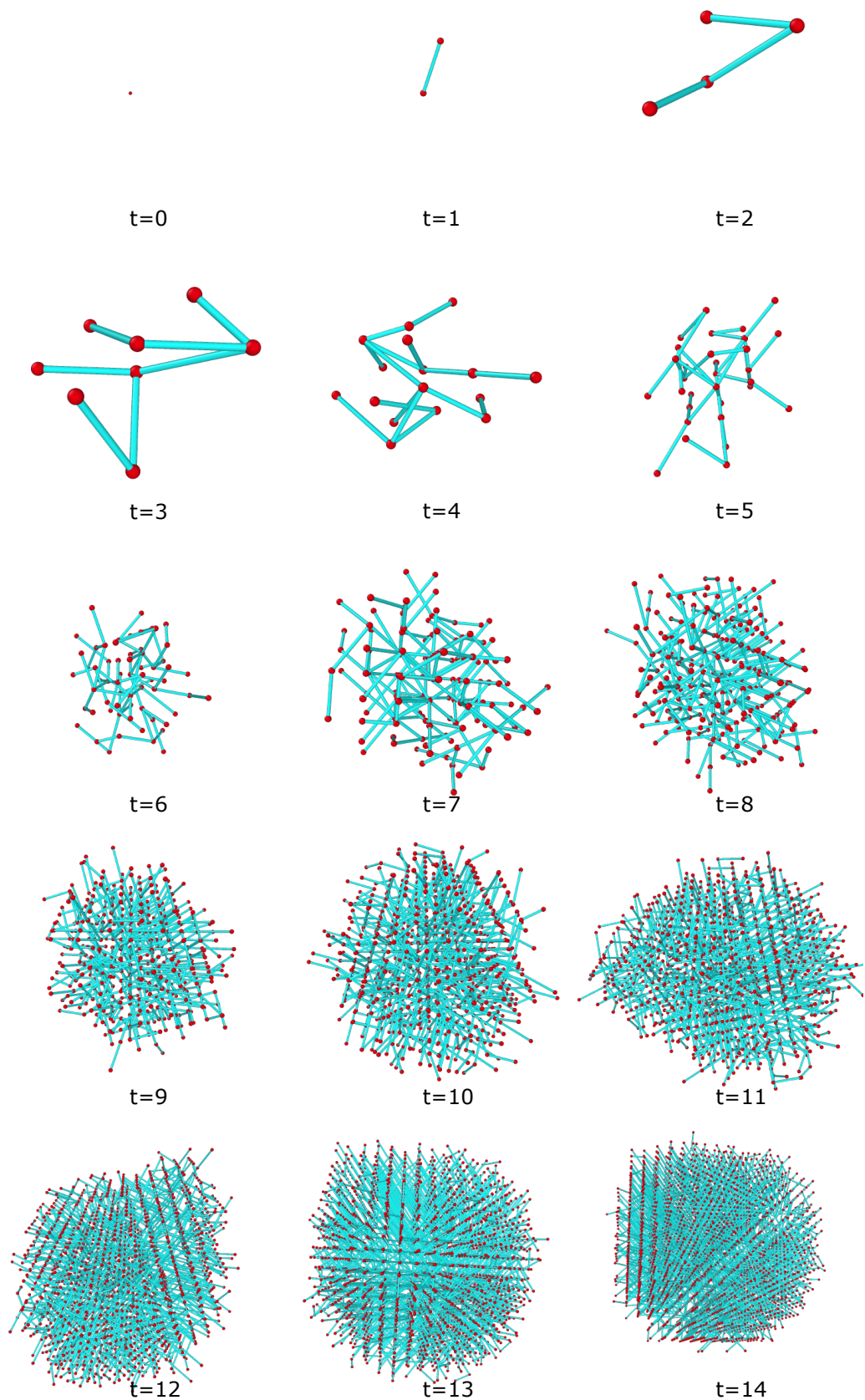}
\caption{{\bf Snapshots of the tumor evolution for $\alpha=1$}. The red nodes are the cells and the blue edges denote the parent-child relationship.}
\label{fig:subsample}
\end{figure}

\clearpage

\bibliography{ith_reference}
\bibliographystyle{unsrt}